\documentclass[sigconf, preprint]{acmart}

\usepackage{booktabs}
\usepackage{graphicx}
\usepackage{epsfig}
\usepackage{subfigure}
\usepackage{balance}
\usepackage{natbib}

\begin{document}
\copyrightyear{2018} 
\acmYear{2018} 
\setcopyright{acmlicensed}
\acmConference[ICMR '18]{2018 International Conference on Multimedia Retrieval}{June 11--14, 2018}{Yokohama, Japan}
\acmBooktitle{ICMR '18: 2018 International Conference on Multimedia Retrieval, June 11--14, 2018, Yokohama, Japan}
\acmPrice{15.00}
\acmDOI{10.1145/3206025.3206053}
\acmISBN{978-1-4503-5046-4/18/06}

\graphicspath{{Figures/}}
\title{Ranking News-Quality Multimedia \textsuperscript{*}}
\thanks{* Please cite the ICRM'18 version of this paper.}

\author{Gon\c{c}alo Marcelino}
\affiliation{%
  \institution{NOVA LINCS}
  \streetaddress{Faculdade de Ci\^{e}ncias e Tecnologia}
  \city{Universidade NOVA de Lisboa} 
  \postcode{2825-516 Caparica}
}
\email{goncalo.bfm@gmail.com}

\author{Ricardo Pinto}
\affiliation{%
 \institution{NOVA LINCS}
  \streetaddress{Faculdade de Ci\^{e}ncias e Tecnologia}
  \city{Universidade NOVA de Lisboa} 
  \postcode{2825-516 Caparica}
}
\email{rjr.pinto@campus.fct.unl.pt}

\author{Jo\~{a}o Magalh\~{a}es}
\affiliation{%
 \institution{NOVA LINCS}
  \streetaddress{Faculdade de Ci\^{e}ncias e Tecnologia}
  \city{Universidade NOVA de Lisboa} 
  \postcode{2825-516 Caparica}
}
\email{jmag@fct.unl.pt}

\renewcommand{\shortauthors}{G. Marcelino et al.}

\begin{abstract}
News editors need to find the photos that best illustrate a news piece and fulfill news-media quality standards, while being pressed to also find the most recent photos of live events. Recently, it became common to use social-media content in the context of news media for its unique value in terms of immediacy and quality. Consequently, the amount of images to be considered and filtered through is now too much to be handled by a person. 
To aid the news editor in this process, we propose a framework designed to deliver high-quality, news-press type photos to the user. The framework, composed of two parts, is based on a ranking algorithm tuned to rank professional media highly and a visual SPAM detection module designed to filter-out low-quality media. The core ranking algorithm is leveraged by aesthetic, social and deep-learning semantic features.
Evaluation showed that the proposed framework is effective at finding high-quality photos (true-positive rate) achieving a retrieval MAP of 64.5\% and a classification precision of 70\%.

\end{abstract}



\maketitle

\section{Introduction}
\begin{figure}[h]
	\centering
	\includegraphics[width=0.45\columnwidth]{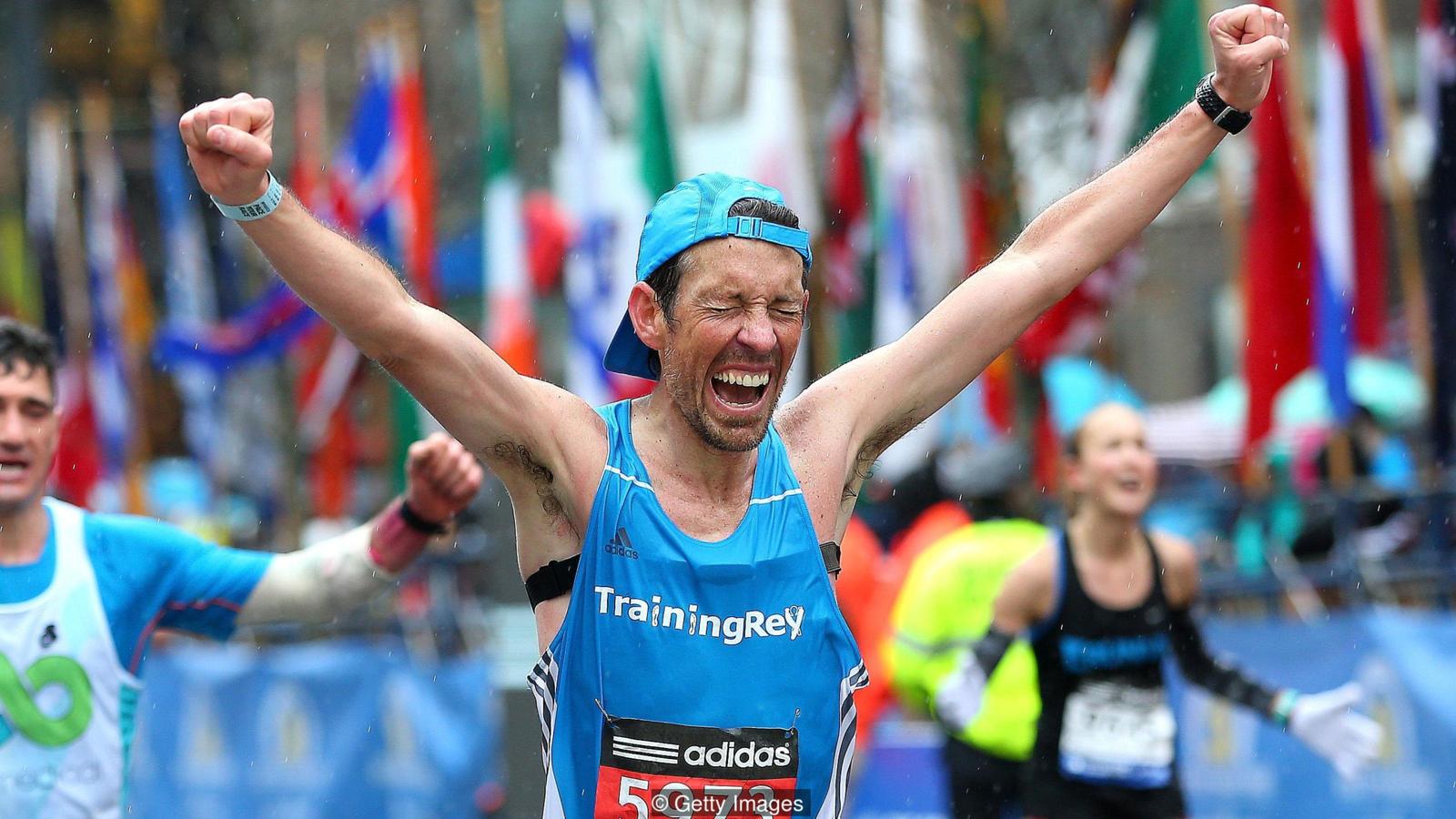}
	\includegraphics[width=0.45\columnwidth]{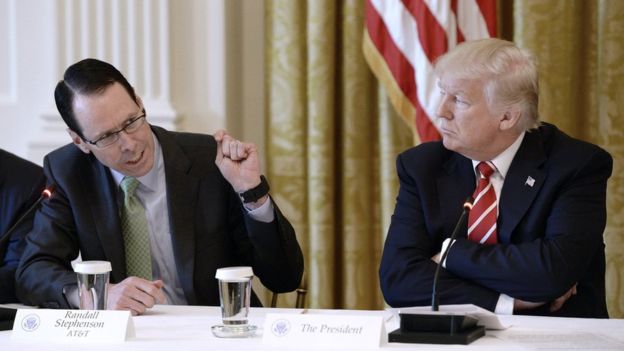}
	\includegraphics[width=0.45\columnwidth]{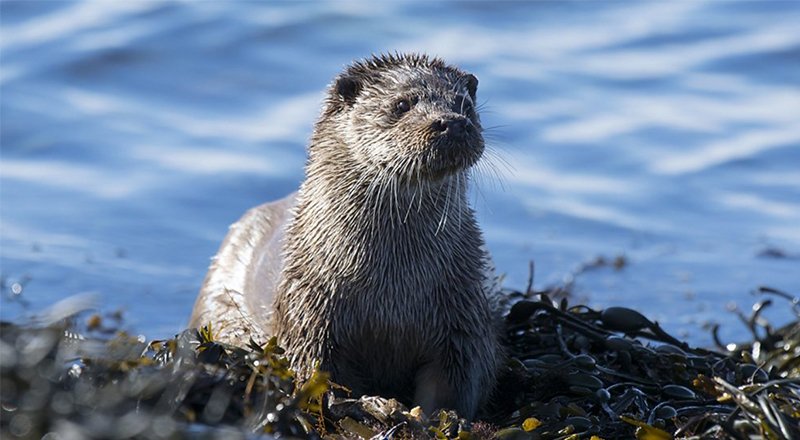}
	\includegraphics[width=0.45\columnwidth]{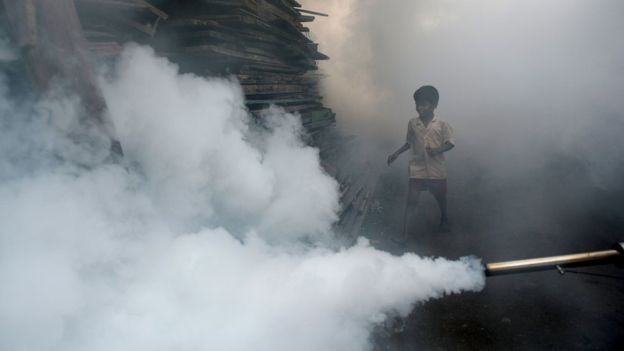}
	\caption{News-quality photos.}
	\label{fig:correct}
\end{figure}

Picking images to be used in the context of news media is a difficult and nuanced task, normally attributed to news editors. 
The process itself is complex and takes into account many variables: the visual quality of the image, how it relates to the news it is supposed to illustrate and how much of the story it conveys by itself, are just some of them \cite{kobre2004photojournalism}. 
However, with the advent of large scale social media platforms like Twitter, Facebook and Flickr, interesting and appealing images that can illustrate a piece of news are no longer only present in the portfolio of news photographers. User generated content has become a great source of news images as the photographic quality of mobile devices continues to rise. Additionally, social media users document the events they participate in themselves, taking photos that cover many more places and perspectives, than a group of news photographers could ever be able to create. Consequently there is now too much content for a news editor to filter through, as the job of crawling Twitter for photographies as the ones in Figure~\ref{fig:correct}, that capture an event while maintaining high visual quality standards, cannot be attributed to a human.

To aid the news editor in this task, we propose a framework designed to filter and rank event related media according to news professionals standards. We aim at reducing and simplifying the task of the news editor, from choosing an image from a set of thousands, including SPAM and low quality images, to choosing an image from a small set of automatically ranked high quality images. 

Consequently, our main contribution is the specification of a machine learning based approach for selecting high quality media that can be used to represent a piece of news.
To model and quantify the photographic quality that news-editors are looking for, our hypothesis is that one needs to consider the problem across three fundamental dimensions: aesthetics, semantic and social. The argument is that ranking by visual aesthetics alone, is not enough -- the sharpness and colorfulness of pictures needs to be complemented by strong and clear semantic content. Also, getting some preliminary human feedback is crucial, hence, social features are also an important element. 

Considering these three domains of a photography (visual, semantic and social) is an important step towards the solution of the problem, but one still lacking in the ability to capture its full complexity. First, we need to consider the non-linear decisions that news editors make when assessing the news-quality of a photo. To overcome this challenge, the proposed framework uses Gradient Boosted Trees~\cite{chen2016xgboost} (GBT) that can capture such non-linear relations. Second, since SPAM is a big part of the content present in social media we explicitly tackle the task of SPAM detection, to ensure that low-quality photos such as memes and adverts are not even considered for analysis. We do this by reviewing the textual and visual components of the content under scrutiny. Enforcing this specialized SPAM detection methodology allows us to simplify the task of the filtering and ranking models, as these can be designed to solely work with content that is beyond a basic threshold of quality. Finally, we remove redundant duplicated content from the list of photos presented to the news-editor, ensuring the non-redundancy of the images suggested.

The rest of the paper is organized as follows: Section 2 describes related works that either influenced this work or that can be used to complement it.
Section 3 elaborates on the structure of the proposed framework. Section 4 describes the methodologies developed to classify and rank images according to their usability in the context of news media. Section 5 deals with the methods and algorithms used for SPAM detection. Finally, Section 6 scrutinizes the evaluation processes the framework was subjected to, while Section 7 concludes the paper.

\section{Related work}

The proposed framework is designed to filter and rank social media content according to news-quality standards. Hence, news illustration is a close problem that addresses some of the same challenges.
\citet{marchesotti2011assessing} propose a system to illustrate events, where semantic inferencing and visual analysis processes are combined to automatically find media to illustrate events, in Internet services and platforms. 
Additionally, \citet{liu2013eventenricher} propose a process for finding media related to an event, by leveraging information about the event against content metadata and additional textual information. As a result, one could use these methodologies to find content to be filtered and ranked by the proposed framework. 

Although the study of methods for measuring the quality of news pieces, such as the one described in \cite{arapakis2016linguistic}, is a fairly popular topic, not much work can be found dedicated to the development of tools to help journalists in their tasks, specially when the task is picking media to illustrate news.
Due to this, one may turn to works tackling other content filtering and ranking tasks for context and information.
\citet{harper2008predictors} propose an architecture for qualifying content in Q\&A dedicated forums like Yahoo! Answers. Although the objectives are different, we take advantage of some of these techniques. Namely we take into account social signals and what is referred to, by the authors, as \textit{intrinsic content quality}, while also making use of the semantic characteristics of the content.

Works on aesthetic assessment were also considered. Although we are not directly interested in finding aesthetic pleasing images, we are aware of the importance of visual quality and aesthetics in news media content. In this context, works such as \cite{dhar2011high} propose the use of high-level features, like compositional attributes, content of the photography and illumination quality, as a way to assess aesthetic quality.
On the other hand, \citet{marchesotti2011assessing} make use of low-level generic features and supervised learning techniques to infer the aesthetic quality of images.
Taking both the importance of low and high level features into account, we considered a large set of image features, a subset of them extracted using the tool also employed in \cite{martins2017semi}, making use of visual features of different, but possibly complementary, levels of abstraction.

Finally, and now relating to SPAM detection, both McParlane et al. \cite{Mcparlane2014} and Schinas et al. \cite{Schinas2015} address the problem of automatically detecting images unsuitable for visual summarization tasks, stating the importance of dealing with images such as "memes" and captioned images. Particularly, in \cite{Mcparlane2014}, the authors also deal with the problem of finding duplicated and near-duplicated images using techniques such as pHash, to avoid presenting the same image to a user twice.  
Since many of the images by us considered as SPAM fall into the "synthetic images" category (such as digital flyers and some "memes") we also searched for previous work regarding the topic. Even though synthetic image taxonomy is not a field that has received a lot of attention so far, \cite{Lienhart2002}, \cite{Mcparlane2014} and \cite{Wang2006} hint at some low level features that help distinguish synthetic images from photographs, such as color histograms, edge histograms and geometric shape counters. 
To increase the performance of our synthetic image classifier, we make use of some of these suggested features, proposing our own in the process.

\begin{figure*}[htbp]
	\centering
	\includegraphics[width=\textwidth]{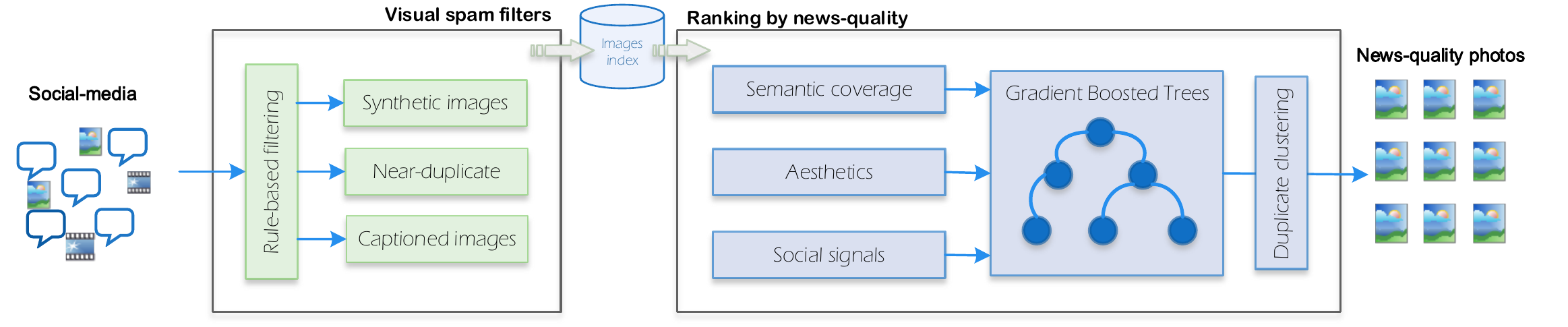}
	\caption{The framework for ranking news-quality pictures in social-media is leveraged by a machine learning algorithm that merges social, visual, semantic and aesthetic evidence.}
	\label{fig:archi}
\end{figure*}

\section{Finding News-Quality Pictures}
Our goal is to find news worthy photos that depict real life events. Therefore, we want unedited photographs (unless the editing aims to improve aesthetic quality), representing the viewpoints of peple witnessing the event. Figure~\ref{fig:archi} illustrates the architecture of the proposed framework. Its main components are:

\begin{itemize}
    \item \textbf{Social-media listener.} We chose Twitter as the source of social-media pictures. This choice is supported by a proven correlation between what is posted on the social network and news media. As an example, \cite{Kwak:2010:TSN:1772690.1772751} shows that over 85\% of the topics trending on Twitter are also covered by the news.
    
    \item \textbf{Visual SPAM filter.} The stream of social media posts are first processed and filtered by a spam detection module. This way, images such as memes, adverts, and images of extremely low resolution, are immediately removed from the pipeline and are never considered in the ranking and classification processes.
    
    \item \textbf{Visual redundancy.} The content that is not discarded by the Visual SPAM filter is then processed by the duplicate and near-duplicate image detection algorithms.
    
    \item \textbf{News-quality ranking.} Finally we have a component charged with ranking photos by their news-quality, i.e., a machine learning model, Gradient Boosted Trees, that combines aesthetic, semantic and social criteria to infer how news worthy an image is.
\end{itemize}

In the following sections, the above components are presented in detail.

\section{Ranking by News-Quality}
Determining if a picture has news-quality is a complex task that cannot be solved by taking only into account its visual appeal. The picture can, for instance, be visual appealing but severely lacking in interesting content and information. To solve this problem we consider not only the \textit{visual aesthetics} of pictures, but also the \textit{semantic content} and the \textit{social signals} associated with them. Moreover, we argue that there are non-linear pairwise interactions among these distinct sets of features. Due to this, and inspired by the work of~\cite{harper2008predictors}, we propose to solve the present problem with Gradient Boosted Trees -- a tree based machine learning model designed for supervised learning. 

Besides being robust to outliers, Gradient Boosted Trees (GBT) are known to work well with categorical and continuous data, which is a critical advantage to solve our problem, where both types of features co-exist. Additionally, GBT also works well with both larger and smaller datasets, which means the ability to train the model with sets of news related media of different sizes. Finally, additional benefits of GBT also include the fact that it performs implicit feature selection, the ability to deal with non-linear relationships in the data as well as capture high-order interactions between features, making it, overall, a very versatile model. These, and other advantages are discussed in greater detail in \cite{friedman2001elements} and \cite{murphy2012machine}.

While we also tested other models, such as Linear and Ridge regression models, $SVM^{rank}$ (an instance of $SVM^{struct}$~ \cite{joachims2006training}), Naive Bayes and Logistic regression, in the end, the model that yielded the best performance was GBT.

\begin{table}[t]
\centering
\caption{Visual features and respective descriptions}
\label{tab:visual_features}
\begin{tabular}{p{1.5cm}p{6cm}}
\toprule
Feature & Description\\
\midrule
\#Edges & the number of vertical, horizontal and diagonal edges present in an image.\\
Rule of $1/3$ & real value representing how much an image complies with the commonly used photography composition rule.\\
Focus & real value describing how focused an image is. \\
Entropy & real value measuring an image's entropy.\\
Faces & the number of human faces present in an image.\\
Luminance & real value describing an image's brightness.  \\
Simplicity & real value representing how simple an image is in therms of the distribution of its colors.\\
Area & the width $\times$ height of an image in pixels.\\
Aspect & the height of an image divided by its width.  \\
Orientation & if an image is square or in a portrait or landscape orientation. \\
Colorfulness & real value describing an image's colorfulness.\\ 
\bottomrule
\end{tabular}
\end{table}

\subsection{Visual quality}
Deciding if a photo is news worthy is a very subjective task. Nevertheless, when approaching news media one expects a certain set of characteristics to be present in its visual content, even if only subconsciously. In order to make use of these latent characteristics we extracted a large set of visual features to quantify the visual quality of photos. This large set of features allows GBT to perform implicit feature selection and capture complex feature interactions. These features are presented and described in Table~\ref{tab:visual_features}, being that the first seven were extracted using the image feature extractor made available\footnote{\url{https://github.com/pcpmartins/extractor}}
by \cite{martins2017semi} and Colorfulness was extracted through the method proposed in \cite{hasler2003measuring}. While the intuition for choosing some of these features might seem obvious, it should be noted that the feature regarding the number of faces present in a photo was chosen as works such as the one presented in \cite{isola2011understanding} show the positive visual impact of the presence of faces in photographies. Additionally, aspect ratio was added as a feature because certain photography equipment is directly associated with a specific image aspect ratio. As an example, DSLR's, cameras normally used in a professional or semi professional setting, normally output images with an aspect ratio of 3:2 \cite{freeman2007photographer}.

\begin{table}[t]
	\caption{Most common concepts associated with news-worthy and non-news-worthy images ordered by decreasing probability of appearance.}
	\label{tab:semantic_prob}
	\centering
	\begin{tabular}{lc|lc}
		\toprule
		Low quality concepts & Prob. & High quality concepts & Prob.\\

		\midrule
        performance & 0.198 & product & 0.282\\
        performing arts & 0.189 & fun & 0.225\\
        event & 0.189 & event & 0.134\\
        entertainment & 0.153 & advertising & 0.126\\
        fun & 0.153 & font & 0.118\\
        performance art & 0.126 & girl & 0.115\\
        recreation & 0.117 & facial hair & 0.099\\
        dancer & 0.108 & recreation & 0.099\\
        crowd & 0.108 & stage & 0.095\\
		product & 0.108 & performance & 0.092\\
        \bottomrule
	\end{tabular}
\end{table}

\subsection{Image concepts}
Our initial intuition, was that images that are used to illustrate news pieces have a particular distribution of concepts associated to them. As Table~\ref{tab:semantic_prob} shows, concepts such as \textit{product} are expected to be less frequent in news media images then a concept like \textit{stage}. Using this knowledge, we propose a way to calculate two additional features that take advantage of these trends to improve our filtering and ranking methodologies.

Given two sets, $Y$ containing images known to have news-quality, and $N$ containing images known to not have news-quality, we first calculate $Py_i$ and $Pn_i$ the probability of concept $i$ appearing in the images present in $Y$ and $N$, respectively. We do this for all concepts extracted from the images in $Y$ and $N$, that appear in more then one image. When a new image containing the set of concepts $C$ is given as input to the framework, both $\sum_{x \in C }^{} Py_x$ and $\sum_{x \in C }^{} Pn_x$ are calculated so that they can be used by the Gradient Boosted Trees model. These values are the sum of probabilities of each concept belonging to an image that is news worthy and not news worthy, respectively. 

Although multiple image concept extraction methodologies are currently available, we choose to use the Google Cloud Vision API for the purpose. We considered a total of 850 unique concepts and, on average, each image was annotated with 7.7 concepts. Table~\ref{tab:semantic_prob} shows the most common concepts associated with the news worthy and non news worthy images present in the social-media photos datset described in Section 6.1.

\subsection{Social signals}
Given the subjectivity associated with the task of identifying news worthy images, it is important to take into account not only the data extracted directly from the images but also the social signals generated by the users who interacted with the associated social media post, from which the images were extracted. 
In practice, we consider the following social signals:
\begin{itemize}
    \item \textbf{\#RT:} number of retweets associated with the post the image was extracted from;
    \item \textbf{\#FL:} the number of followers associated with the user who posted the tweet containing the image;
    \item \textbf{\#UN:} the number of times the image is featured in individual posts;
    \item \textbf{\#DD:} the number of times a visually near-duplicated image is featured in individual posts.
\end{itemize}

This information is used as a proxy for the users opinions regarding an image's importance and its entertainment and informative value.

\section{Visual SPAM and redundancy}
\label{sec:spam_filter}
\begin{figure}[t]
	\centering
	\includegraphics[width=0.41\columnwidth]{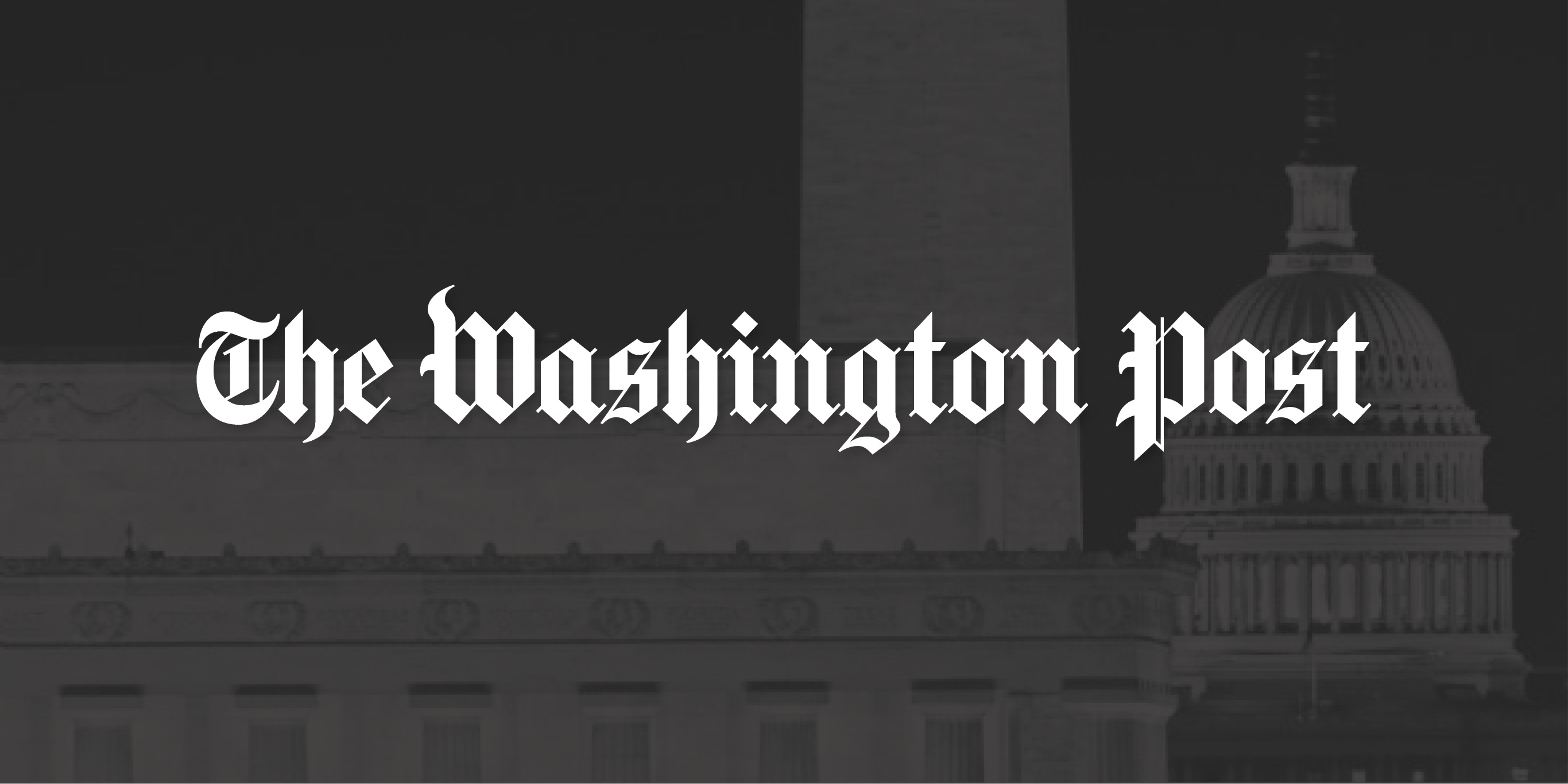}
    \includegraphics[width=0.34\linewidth]{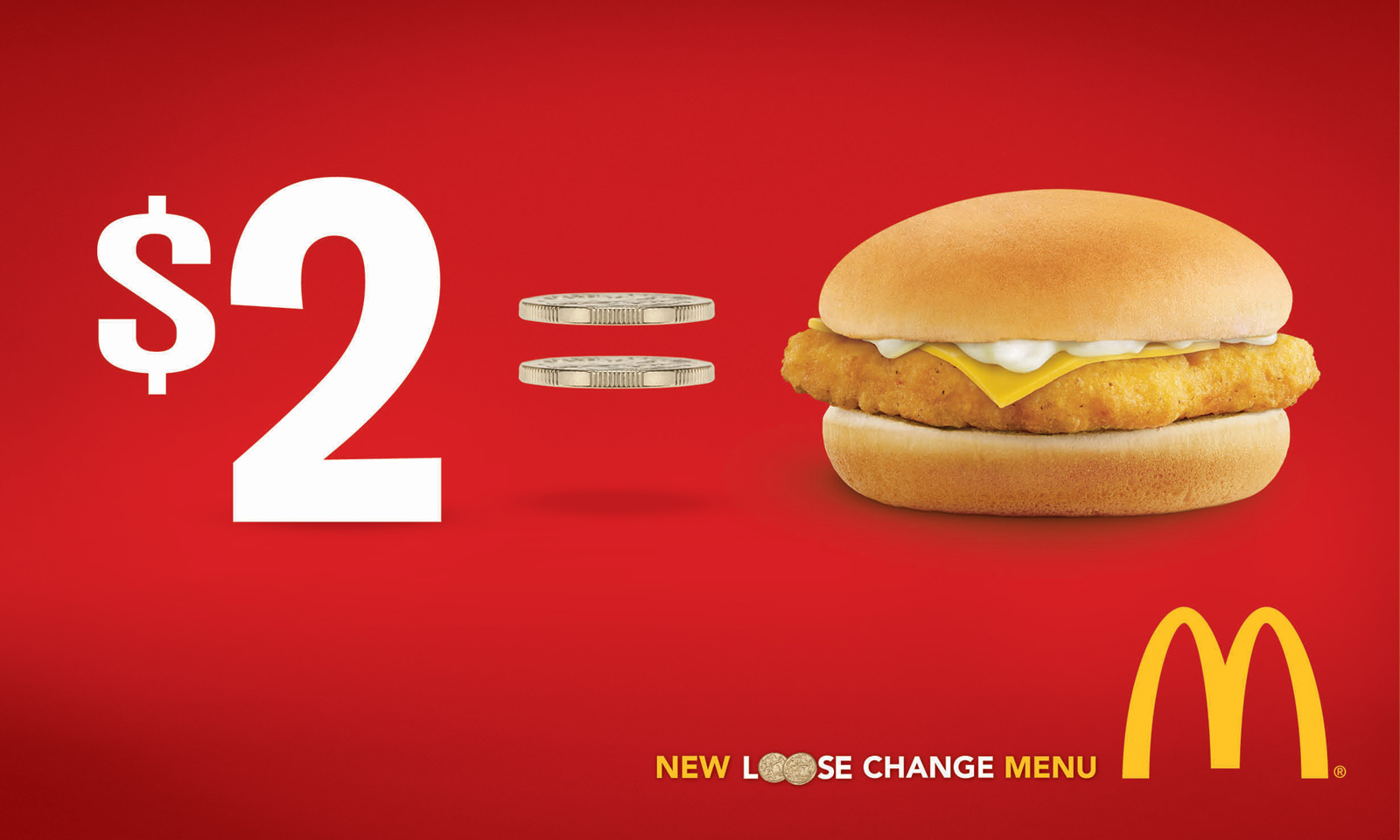}
    \includegraphics[width=0.41\linewidth]{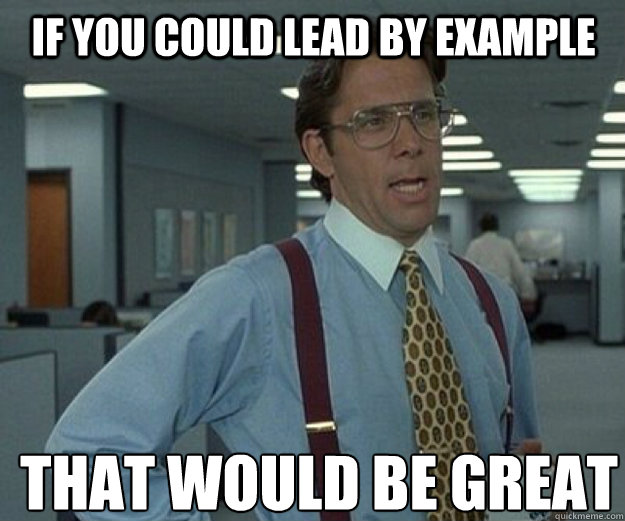}
	\includegraphics[width=0.34\columnwidth]{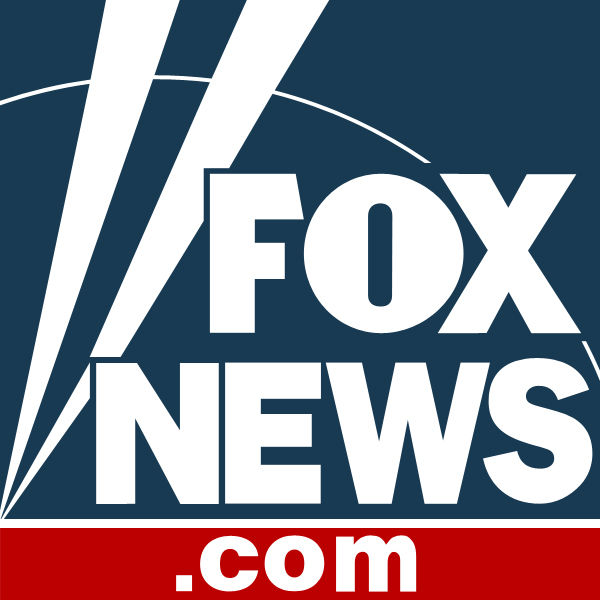}
	\caption{Examples of unwanted images that can be immediately discarded (i.e., logos, adverts and memes).}
	\label{fig:spam}
\end{figure}

Images of adverts, captioned images, memes and similar visual content are big portion of the content posted by social-media users which must be filtered by the framework. To solve this problem we propose a method to filter low-quality and redundant visual information, in order to prevent content like the one presented in Figure~\ref{fig:spam} from being indexed together with valid photos.

We propose a filtering pipeline composed of four distinct parts. The first is the application of a set of simple thresholds well established in literature \cite{mcminn2013building} to features extracted from both the images and posts they were taken from. The second is the usage of a linear model trained to detect synthetic images, to filter images such as digital adverts. The third is the application of OCR technology to subsequently filter out captioned images, such as memes.
Finally, the forth part deals with the large number of duplicated images found among social-media content. The pipeline is detailed in the following subsections.

\subsection{Coarse filtering}
To filter thumbnails, banners and adverts, we follow the approaches of \citet{Mcparlane2014} and \citet{mcminn2013building}, and exclude images extracted from posts that contain more than 3 hashtags, more than 3 mentions or more than 2 URLs. Additionally we also discard small images that, due to their size, are not useful in a news context (i.e., images with less than 200 pixels of width or height).

\subsection{Synthetic images detection}
Several works on synthetic image detection \cite{Lienhart2002, Wang2006}, have shown that some simple low-level features allow to discriminate synthetic images from photographies with great success. We trained a logistic regression model to distinguish synthetic images from real photos using some of the features proposed by \cite{Lienhart2002, Wang2006}: \textit{number of corners}, \textit{number of vertical and horizontal lines}, \textit{number of dominant colors}, \textit{most common color (C1)}, and 3 features derived from the color transitions feature (the measure of color distance between two neighbor pixels).

\subsubsection{Color transitions}
Color transitions \cite{athitsos1997distinguishing} are a common property of real world images. Synthetic images have large portions with very similar colors and limited transitions. Color transitions are calculated as follows: for each pixel $px$ in the image, we compute the color distance in the RGB space, to all its nearest neighbours as $$d_{color}(px) = \sum_{n=1}^{8}(|r_{px}-r_n|+|g_{px}-g_n|+|b_{px}-b_n|)$$ 

From this metric, we are able to extract 3 additional features: the fraction $f_1$ of pixels with a color distance greater than zero, the fraction $f_2$ of pixels with a color distance greater than $1/4$ of the maximum distance, and the ratio between $f_2$ and $f_1$. 
Synthetic images are expected to have a lower value for fraction $f_1$, as they usually contain large regions with the same exact color. On the other hand, photographic images are expected to have a lower value for fraction $f_2$ for the same reason stated in section \ref{sec:feat_most_common_color}, regarding color gradients.

\subsubsection{Most common color}
\label{sec:feat_most_common_color}
It is expected that, in synthetic images, the most common (exact) color will be more frequent across the entire image, whereas photographies will tend to have their colors divided into slightly different hue, saturation or brightness values, forming a color gradient across regions of the image. 
Therefore, we work with \textsl{the ratio between the number of pixels with the most common color and the total number of pixels in the image} as one of the features used to distinguish both classes of images. 

\subsubsection{Dominant colors}
We hypothesize that photographic images in general have a larger number of \textit{dominant colors} than synthetic images.
To calculate this feature, we build a HSV color histogram, subdividing each color space into 8 bins, making up a total of $8^3 = 512$ bins. We then count how many of these bins have a frequency higher than a given threshold. This count corresponds to the \textsl{number of dominant colors} feature.

The threshold was determined by iterating different values, choosing the one that minimizes the normalized difference $Norm_{\Delta_{dom}}$ in the following equations:

\begin{equation}
    \Delta_{dom} = (\mu_{photo}-\sigma_{photo}) - (\mu_{synth}+\sigma_{synth})
\end{equation}

\begin{equation}
Norm_{\Delta_{dom}} = \frac{|\Delta_{dom} |}{\mu_{photo}} + \frac{|\Delta_{dom} |}{\mu_{synth}}
\end{equation}
where $\mu_{photo}$ and $\mu_{synth}$ denote the average number of dominant colors in the images belonging to the \textit{photographic} and \textit{synthetic} classes, respectively and, $\sigma_{photo}$ and $\sigma_{synth}$, their standard deviations.

\subsubsection{Horizontal and vertical lines}
Following the assumption that synthetic images in general have a lot of lines parallel to the frame of the image, we extracted horizontal and vertical lines with the Hough Line Transform, using a minimum threshold of 20 points per line.

\subsubsection{Number of corners}
The number of corners of an image was extracted with the Harris corner detector. 
After applying a threshold to the corner filter response image, one obtains the corners of a given image. To find the optimal threshold we used a synthetic images dataset. We calculated this threshold as a fraction of the maximum score in the corner filter response image and attempted to find the percentage for which the \textsl{ratio between the mean number of corners in the photo class and the mean number of corners in the synthetic class} was the lowest.

\subsection{Captioned images}
Images whose main subject is text tend to be unusable in the context of news illustration. Examples of these types of images include highly captioned images like memes. To filter this type of content we apply a simple prepossessing method to all images and then, using Tesseract \cite{smith2007overview}, an Optical Character Recognition Engine, exclude images containing prominent text. Since the preprocessing method involves applying median blur to the images, and since Tesseract is only able to recognize easily identifiable text, images with small text or text present in the background are correctly left unfiltered by the framework.

\subsection{Visual redundancy}
Since a lot of images present in social-media are slightly altered versions of their respective originals, we propose a way to find not only duplicated but also near-duplicated images. This is important as it means the ability to filter redundant content before it is presented to the news editor.

\begin{figure}[t]
\centering
\begin{subfigure}
  \centering
  \includegraphics[height=0.25\columnwidth]{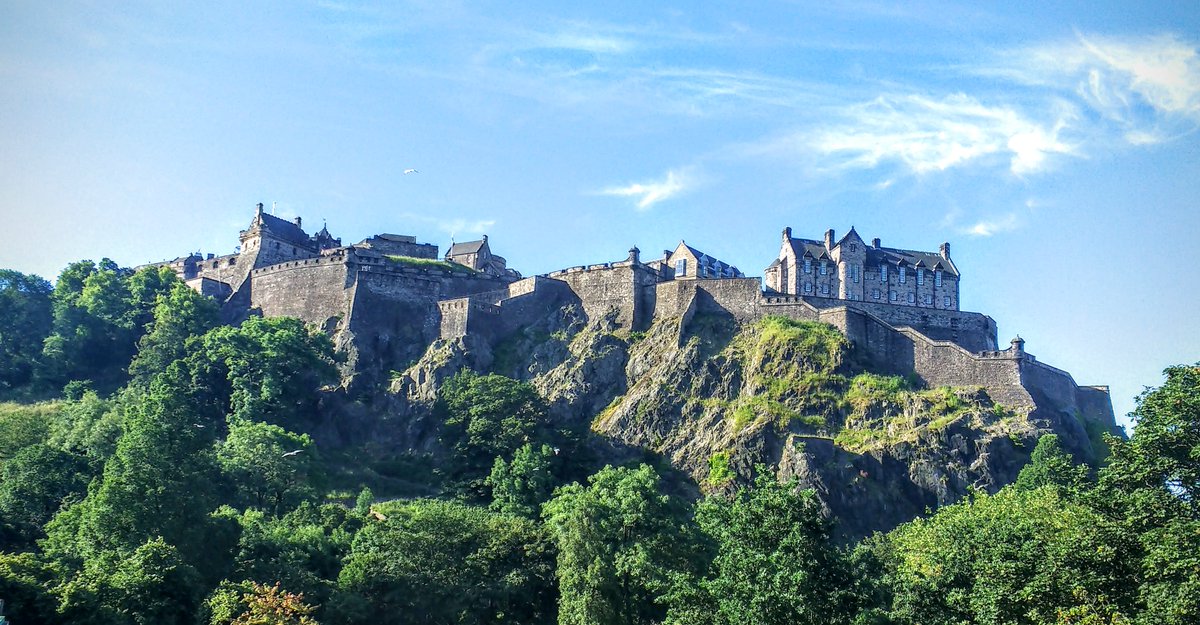}
\end{subfigure}%
\begin{subfigure}
  \centering
  \includegraphics[height=0.25\columnwidth]{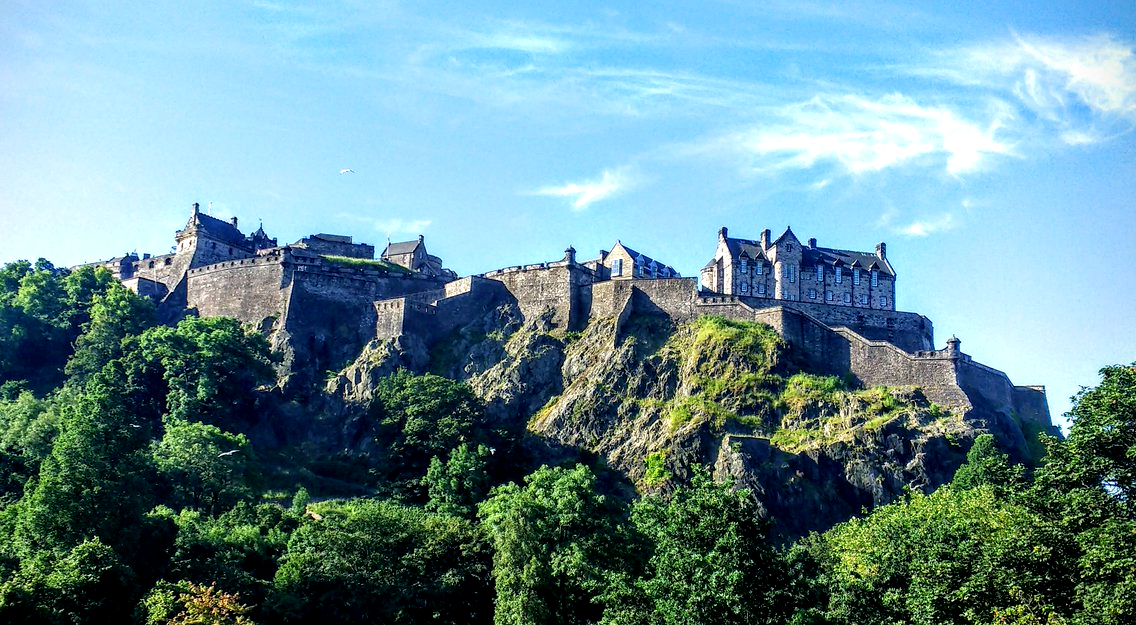}
\end{subfigure}
\caption{Example of near-duplicate images. The first is the original image. The second is a cropped version of the first with different contrast.}
\label{fig:near_duplicates}
\end{figure}

\subsubsection{Duplicate detection}
To access whether two images are exact duplicates of each other we make use of the MD5 hash algorithm. More specifically, we consider two images to be exact duplicates if their respective MD5 hash is the same. Before presenting the results of the ranking and filtering models to the user, all duplicated images are removed.

\subsubsection{Near-duplicate detection} To detect near-duplicated images, we employ perceptual hash (pHash)\footnote{http://www.phash.org/} which relies on the lower frequencies of images, ignoring higher frequencies. This algorithm is a simple and fast method for comparing images, hence our choice. Previous work already proved it presents a high performance in the this task \cite{Tang2012}, regardless if the image is rotated, resized, cropped, exposure compensated or even if small elements are added to it (like a logo or signature). 

As a method for assessing if two images are near-duplicates, we calculate the Hamming distance between their pHash codes, which corresponds to the amount of bit positions where those codes differ \cite{HackerFactor_pHash}. For example, if two images represented by 64 bit codes, are separated by a Hamming distance of 5, it means their pHash codes differ in 5 bits. 
We consider two images to be near-duplicates if the Hamming distance between their pHash values is below 8 as proposed in \cite{Mcparlane2014}. An example of two near-duplicate images can be found in Figure~\ref{fig:near_duplicates}.

\section{Evaluation}

\subsection{Datasets}
To evaluate the different components of the proposed framework, we used three datasets\footnote{Links and features of the used photos are available at \url{http://novasearch.org/datasets/}}.: (i) social-media content from which we need to retrieve high-quality photos; (ii) newswire photos, used to build the high-quality photos model; and (iii) a visual SPAM dataset to learn to filter low-quality photos.

\textbf{Social-media photos.}
We crawled a dataset of 15,439 images from Twitter concerning several real-world concerts and shows that spanned one full month. 
In order to evaluate the ranking algorithm, we created a small sample of 1,500 photos for results pooling. Ground-truth was obtained through crowdsourcing by resorting to 5 annotators that judged the top-\textit{k} photos of each baseline.

\textbf{News-quality photos.}
To create a robust model that is able to qualify photos acording to their news-quality, we obtained newswire photos from The New York Times and the BBC web sites. We collected a total of 100 newswire photos and added 400 social-media photos, sampled from the previous 15,439 images dataset. This dataset comprises a total of 500 images that were annotated by 7 annotators with respect to their "\textit{news-quality provenance}", as described in the following section. Moreover, the annotation effort allowed us to better understand the specific characteristics of news-quality photos.

\textbf{Visual SPAM.}
The detection of visual SPAM requires the detection of different types of images. The NPIC image dataset \cite{Wang2006} contains several photographic and synthetic images, including logos, cartoons and CG images. However, NIPC also contains images that are not common in social-media (e.g. oil paintings) and misses images that are common in social-media (e.g. memes and adverts). To more accurately reflect the type of content found on social-media, we combined some of the NIPC images with social-media images. Hence, we created the new NIPC-Twitter dataset consisting of 13,668 images (638 real-world photos and 390 synthetic images from Twitter, and 10,271 real-world photos and 2,369 synthetic images from NPIC) tailored to the detection of visual SPAM in social-media.

\begin{table}[t]
	\caption{Results of the annotations performed on the news-quality photos dataset according to the question "Could this image have appeared in the New York Times?".}
	\label{tab:agreement}
	\centering
	\begin{tabular}{cccc}
		\toprule
		Agreem. & Images & High quality & LQ/HQ ratio\\
		\midrule
        57\% & 124 & 58 & 1.14 \\
        71\% & 129 & 55 & 1.35 \\
        86\% & 144 & 39 & 2.69 \\
        100\%& 103 & 17 & 5.06 \\
        \midrule
        78\%& 500 & 169 & 1.96\\
        \bottomrule
	\end{tabular}
\end{table} 

\subsection{News-quality photos ground-truth}
Regarding the 500 image dataset used to train the models, all 500 images were annotated via crowdsourcing by 7 annotators. The annotators were presented with the images and asked the question "\textit{Could this image have appeared in the New York Times?}". Table~\ref{tab:agreement} presents, in an abbreviated manner, the results of the annotation process.
Through their answers we can infer the ambiguity of the task, as the 7 annotators only fully agreed on 102 images.
As ground truth for the ranking task, 7 quality levels were attributed to each image according to the number of annotators that voted that the image might have appeared in the New York Times. For this task all 500 images were considered and the regression models used in it were trained to match these quality levels.
As a ground-truth for the classification task only images where 71\% or more of the annotators agreed, were considered. In this case, the image was regarded as having news-quality if the majority of the crowd answered yes to the already mentioned question. The classification models used in the filtering task were trained to match this binary judgment.

As Table~\ref{tab:agreement} shows, out of all 500 images, only 17 of them were classified as possibly having appeared in the New York Times, by all 7 annotators. Of these 17 images, 14 belong to the set of images extracted from news sources, which shows the ability of the annotators to distinguish news-quality images. In Figure~\ref{fig:correct} we present four of these images as examples.

\subsection{Results and discussion}

\subsubsection{Modeling news-quality photos}
The filtering and ranking models were trained using 70\% of the dataset while the results presented next regarding the classification task were measured using the remaining 30\%. We tested models where visual ($GBT_V$), social ($GBT_S$) and semantic ($GBT_C)$ features were used separately and combined ($GBT_F)$ to understand the impact of the different feature sets.
Table~\ref{tab:first_tests} shows the results of these tests. The advantage of joining multiple groups of features, to tackle the proposed task, is being able to attain clearly higher precision, nDCG and MAP values in comparison to the models where only one feature group is used. Consequently using only the visual quality, semantics, or social signals associated with an image as criteria for deciding if it has news-quality, equates to having a worst performance in the task overall. 

\begin{table}[t]
\caption{News-quality filtering task results.}
\label{tab:first_tests}
\centering
\begin{tabular}{l|cc}
	\toprule
	\multicolumn{1}{l}{Features} & Prec. & Acc. \\	
	\midrule
    $GBT_V$   & 0.672 & 0.787 \\
    $GBT_C$   & 0.555 & 0.742 \\
    $GBT_S$   & 0.639 & 0.834 \\
    $GBT_F$   & 0.701 & 0.854 \\
	\bottomrule
\end{tabular}
	\vspace{-5mm}
\end{table}

\subsubsection{Ranking high-quality social-media photos}
The performance of the ranking model was evaluated in an out-of-domain context using the described annotated dataset, created through results pooling.
To do so, we first trained 4 distinct models, the first three taking only advantage of visual ($GBT_V$), social ($GBT_S$) and semantic ($GBT_C$) features individually and the forth using all of the three feature sets simultaneously ($GBT_F$). Each model was then applied to the dataset and the $k$ better ranked images were extracted. These images were, in turn, annotated by 5 annotators again according to the question "\textit{Could this image have appeared in the New York Times?}". Finally, images were classified as news worthy if the majority of the annotators answered yes to the question.
 
\begin{figure}[t]
	\centering
	\includegraphics[width=0.8\columnwidth]{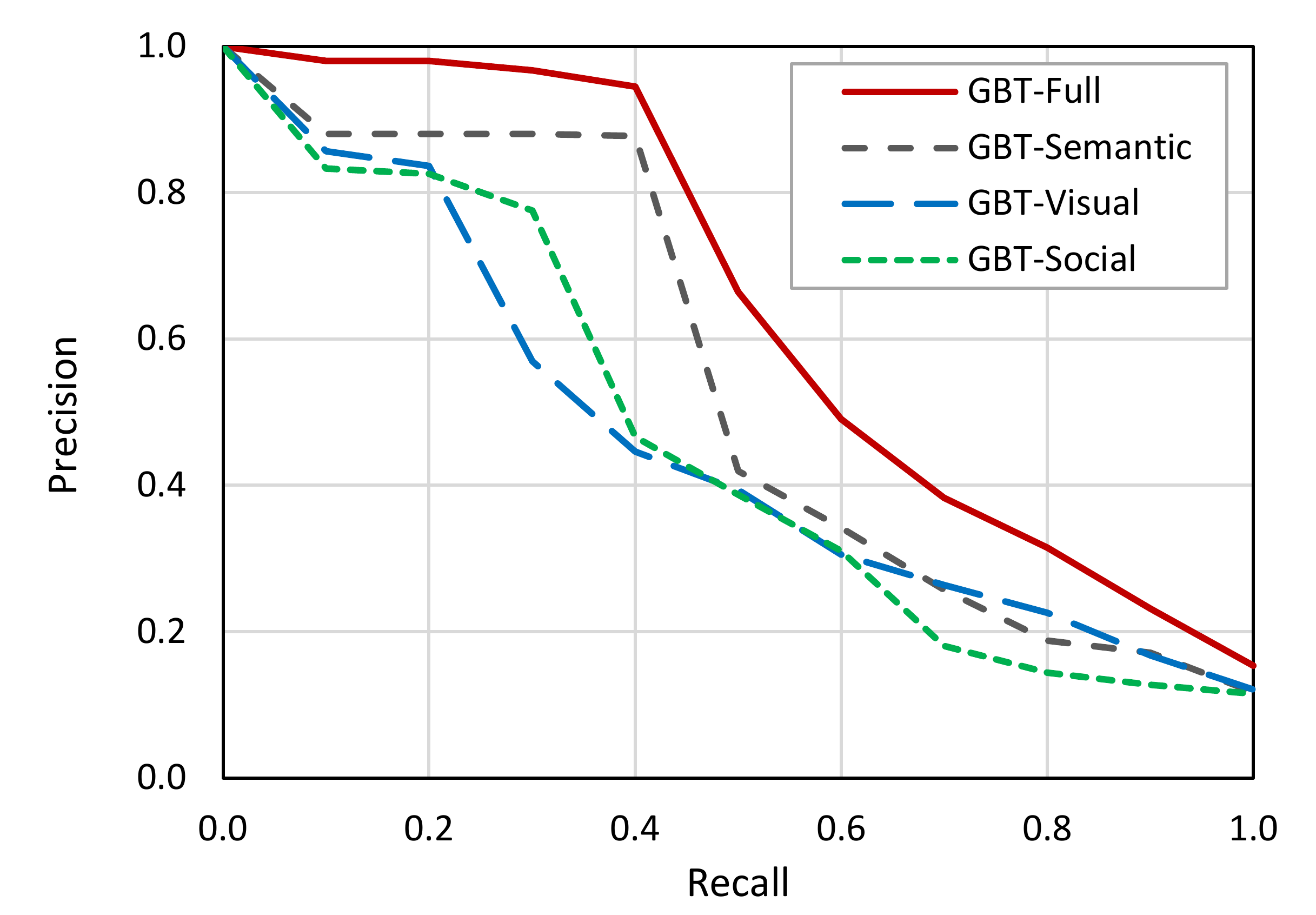}
	\caption{Precision recall curves of the various ranking models.}
	\label{fig:precision_recall}
\end{figure}

 Table~\ref{tab:second_test} shows the precision and nDCG values of the various models tested, while Figure~\ref{fig:precision_recall} presents their precision-recall curve. By analyzing both these metrics we discover that, overall, the models that performed worst were $GBT_V$ and $GBT_S$. In turn the model trained only with semantic features, $GBT_C$, was marginally more successful, specially when retrieving the first half of the relevant images. This shows the importance of semantics in the context of news media. Finally, the complete model ($GBT_F$) was the one that performed better as it was able to take advantage of the combined strengths of the feature groups used.
 
\begin{table}[t]
	\caption{Results of the performance tests done on the various ranking models.}
	\label{tab:second_test}
	\centering
	\begin{tabular}{l|ccc}
		\toprule
		\multicolumn{1}{c}{Features} & Prec@30 & nDCG@50 & MAP\\	
		\midrule
        $GBT_V$ & 0.833 & 0.837 & 0.448\\
        $GBT_C$ & 0.833 & 0.859 & 0.532\\
        $GBT_S$ & 0.733 & 0.836 & 0.454\\
        $GBT_F$ & 0.967 & 0.906 & 0.645\\
		\bottomrule
	\end{tabular}
\end{table}

In Table~\ref{fig:results_example} we exemplify this tendency by examining specific examples of images ranked by each model while identifying, in a broad way, the features that influenced the model's choices. The images $GBT_V$ ranked higher (shown on the left side of the table) are of high visual quality, but the model is unable to ensure the interestingness of the images it presents, the image with the mobile phone being a good example of this problem. $GBT_C$ is able to correctly assert that a photo of a concert is more likely to be used in news media then a \text{selfie}. However the model ends up ranking an extremely blurry image as one of the best in the set, when possibly better suited alternatives were available, like the one displayed on its right. In turn, $GBT_S$ ranks images according to social signals, consequently discarding good images that did not gain social traction. The model ranks correctly images that have a lot of social traction but, when this ceases to be the case, the existing social signals stop being enough to distinguish between images. This tendency is not only observable in Table~\ref{fig:results_example} but also in the precision-recall curve, as the model is the worst for recall values higher then 0.6. Lastly  $GBT_F$ leverages the benefits of the other models to correct, to a degree, their individual faults. The $GBT_F$ model is still able to distinguish a \textit{selfie} from a photo of a concert but is also able to assure the visual quality of the better ranked images. Additionally the model does not focus singularly on social signals meaning that, although these are considered, an unpopular but visual appealing image, semantically tied to news media, is still ranked high by the model.

\begin{figure}[t]
	\centering
	\includegraphics[width=0.9\columnwidth]{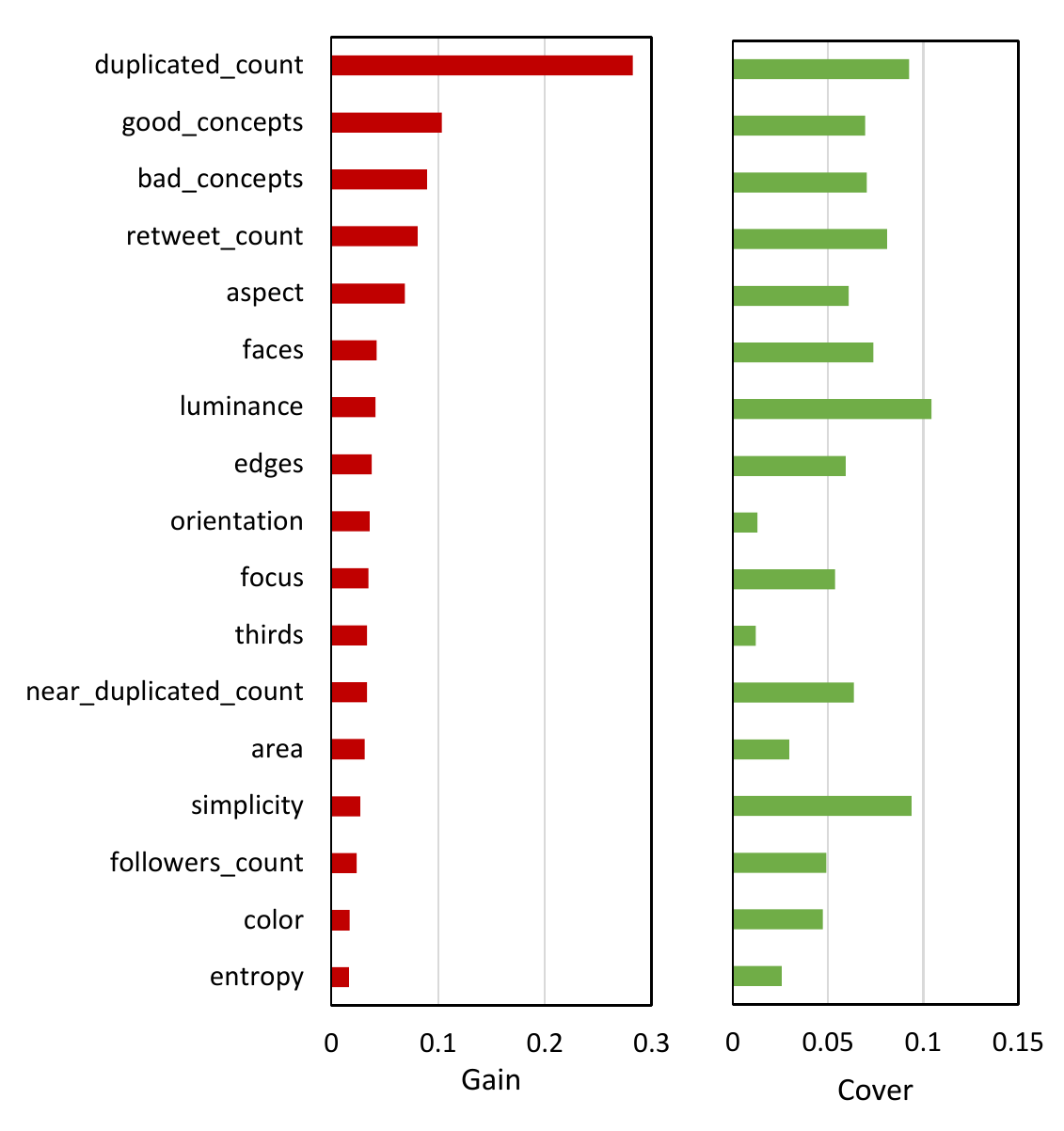}
	\caption{The importance of each feature measured through its gain and cover in the Gradient Boosted Trees regression model.}
	\label{fig:feature_importance}
\end{figure}

Finally, Figure~\ref{fig:feature_importance} presents, for each visual, social and semantic feature, its associated gain and cover in the context of the $GBT_F$ model. The higher the gain, the more important a feature is in improving the accuracy of the model. Similarly, cover equates to the amount of coverage of a feature when used in the trees. Here, the gain table shows that, although most visual features have a small gain individually, the model comprised only of visual features still retains a decent performance due to the high number of distinct visual features used. Additionally we can find visual, social and semantic features in the top 5 features with more gain, confirming that all feature groups increase, by themselves, the performance of the model.
\begin{table}[t]
	\caption{Performance of the photo/synthetic image classifier for several combinations of features.}
	\label{tab:synth_classifier_performance}
	\centering
	\begin{tabular}{l|ccc}
		\toprule
		\multicolumn{1}{l}{Best single features (NIPC)} & Precision  &  Recall & F-measure \\
		\midrule
		Luminance 	& 0.72   &   0.68    & 0.66  \\
		Dom. colors 	& 0.79  & 0.79 & 0.79 \\
		Ratio C1	& 0.85 & 0.81  & 0.80	 \\
		\midrule
		\multicolumn{1}{l}{Best feature set} & Precision  &  Recall & F-measure \\
		\midrule
		NIPC trained	& 0.97  &   0.97 & 0.97 \\
		NIPC-Twitter trained & 0.91 & 0.91  & 0.91\\
		\bottomrule
	\end{tabular}
\end{table}

\begin{table*}[t]
	\centering
	\begin{tabular}{@{}l|cccc@{}}
		\toprule
		\small{\textbf{$GBT_{V}$}}
		& \small{Luminance$^\uparrow$, Focus$^\uparrow$, Color$^\uparrow$}
		& \small{Luminance$^\uparrow$, Focus$^\uparrow$, }
		& \small{Luminance$^\downarrow$, Focus$^\downarrow$}
		& \small{Aspect$^\downarrow$, Faces$^\uparrow$ Focus$^\downarrow$ Entropy$^\downarrow$}\\
	    & \includegraphics[height=0.20\columnwidth]{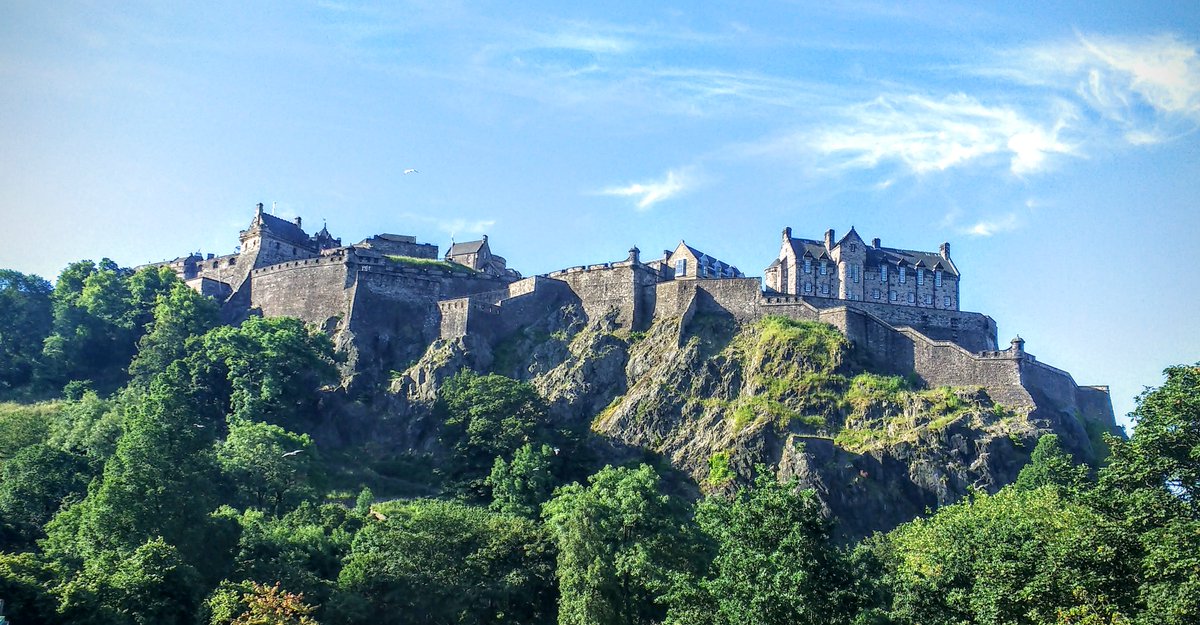}
	    & \includegraphics[height=0.20\columnwidth]{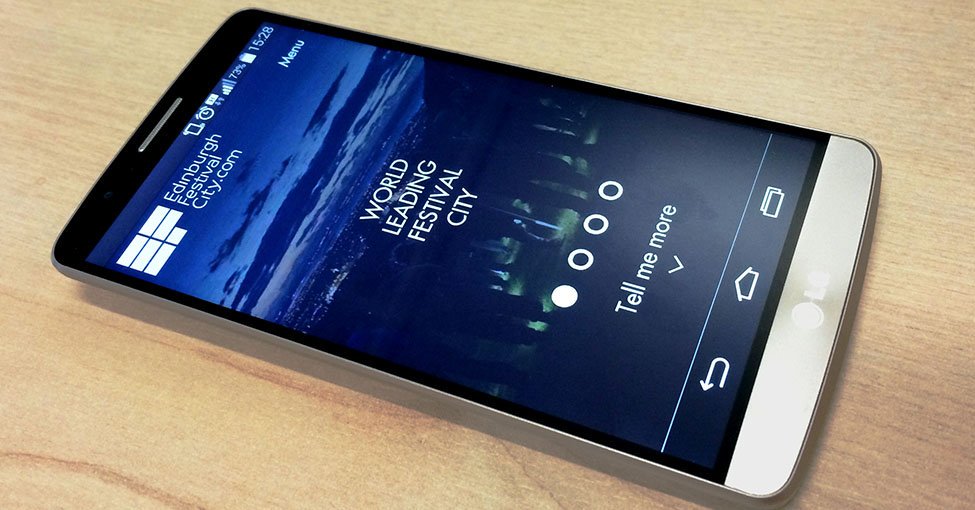}
	    & \includegraphics[height=0.20\columnwidth]{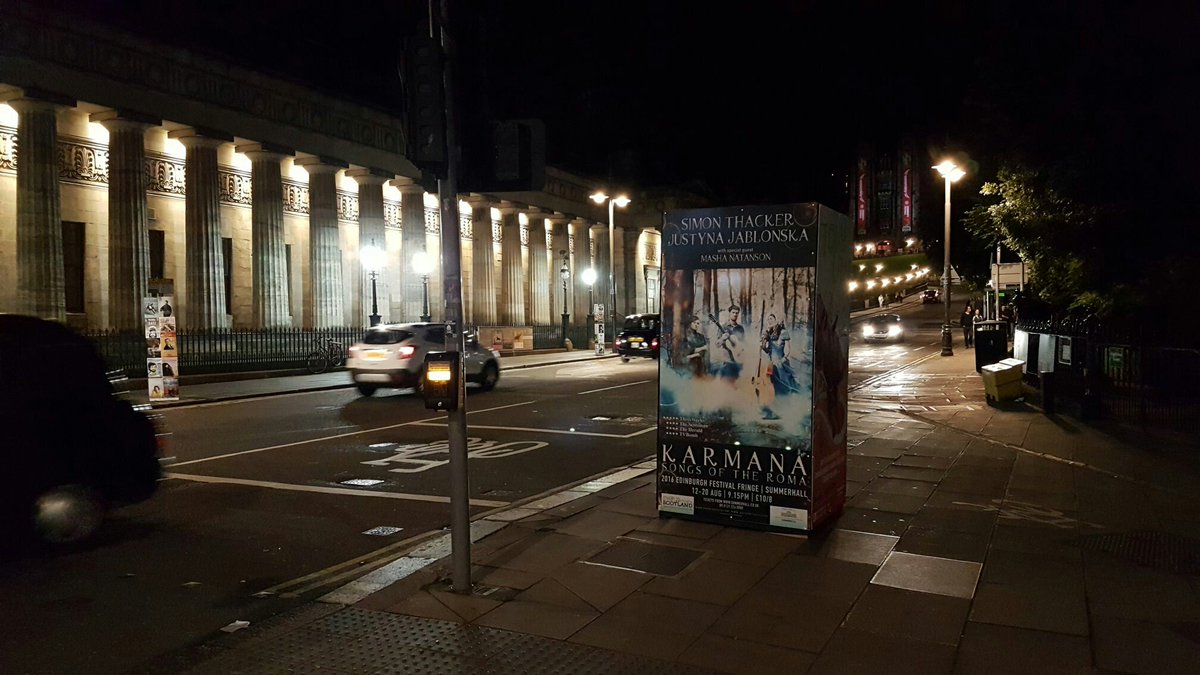}
		& \includegraphics[height=0.20\columnwidth]{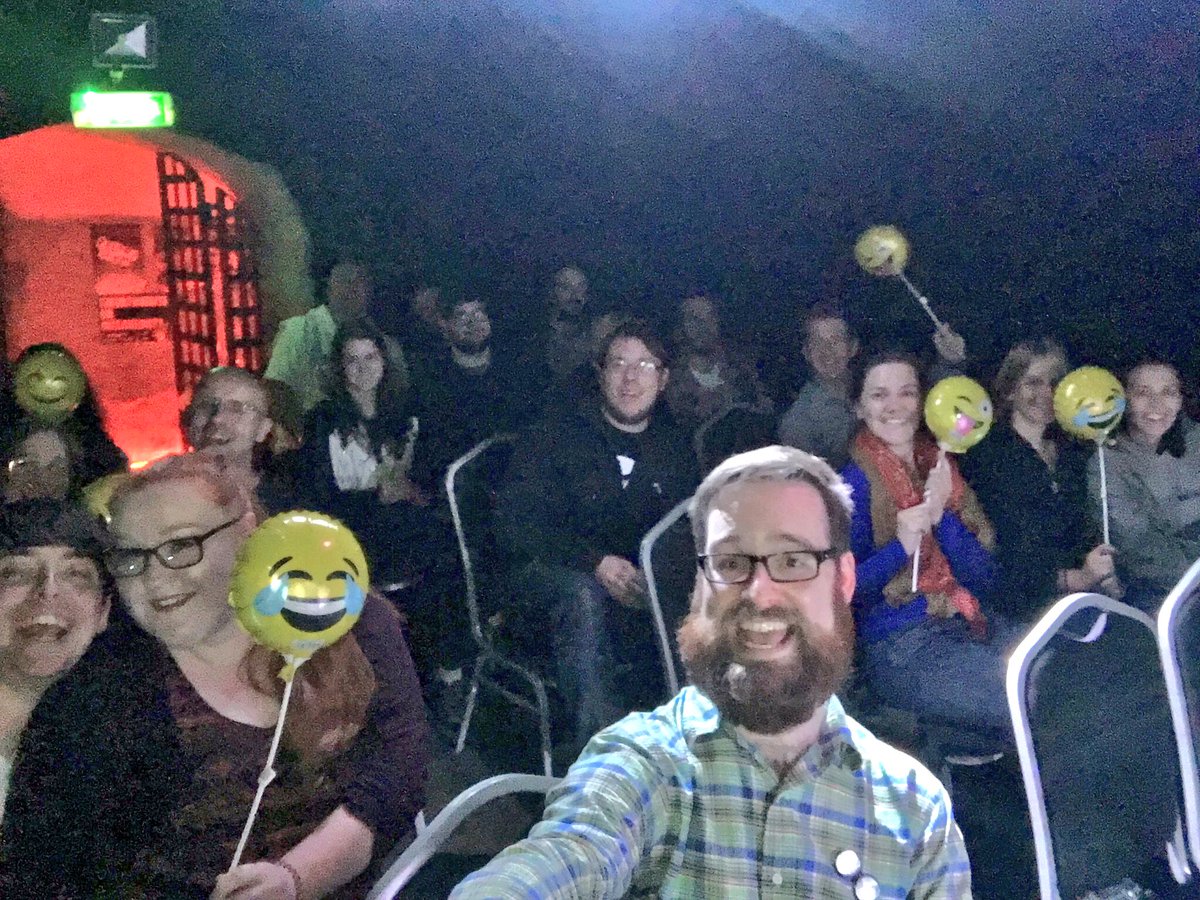}\\
        \midrule
		\small{\textbf{$GBT_{C}$}} 
		& \small{Performing Arts$^\uparrow$, Event$^\uparrow$, Stage$^\uparrow$}
		& \small{Event$^\uparrow$, Festival$^\uparrow$}
		& \small{(No interesting concepts)}
		& \small{Girl$^\downarrow$, Selfie$^\downarrow$}\\
	    & \includegraphics[height=0.20\columnwidth]{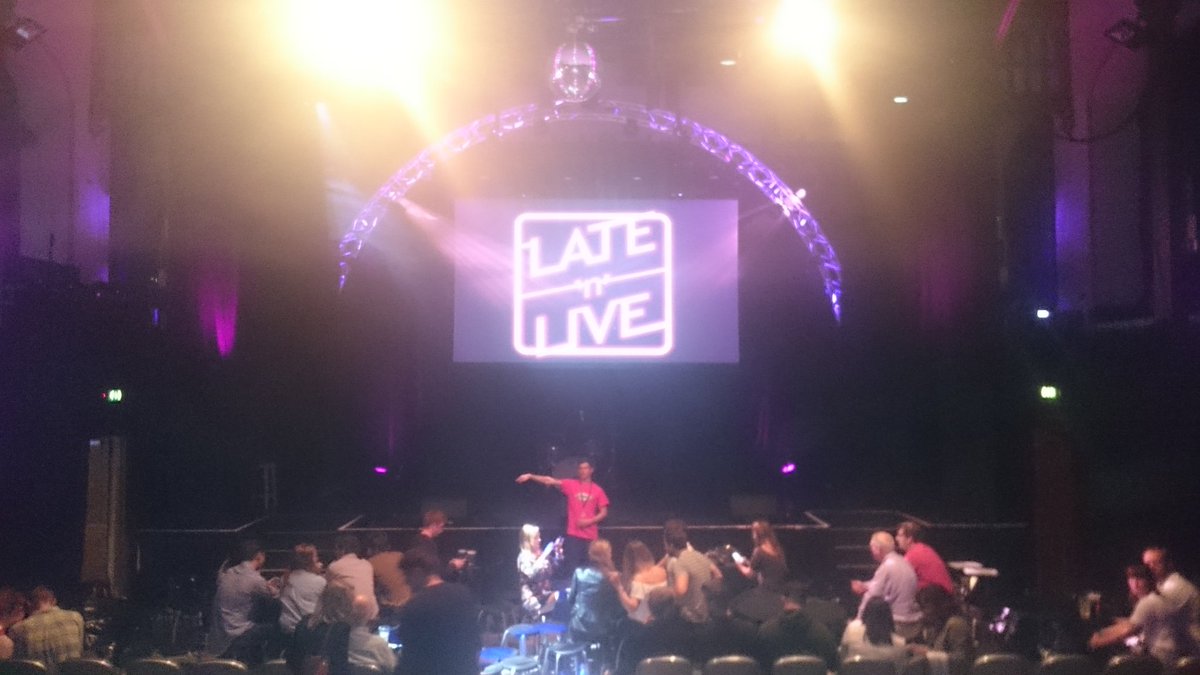}
	    & \includegraphics[height=0.20\columnwidth]{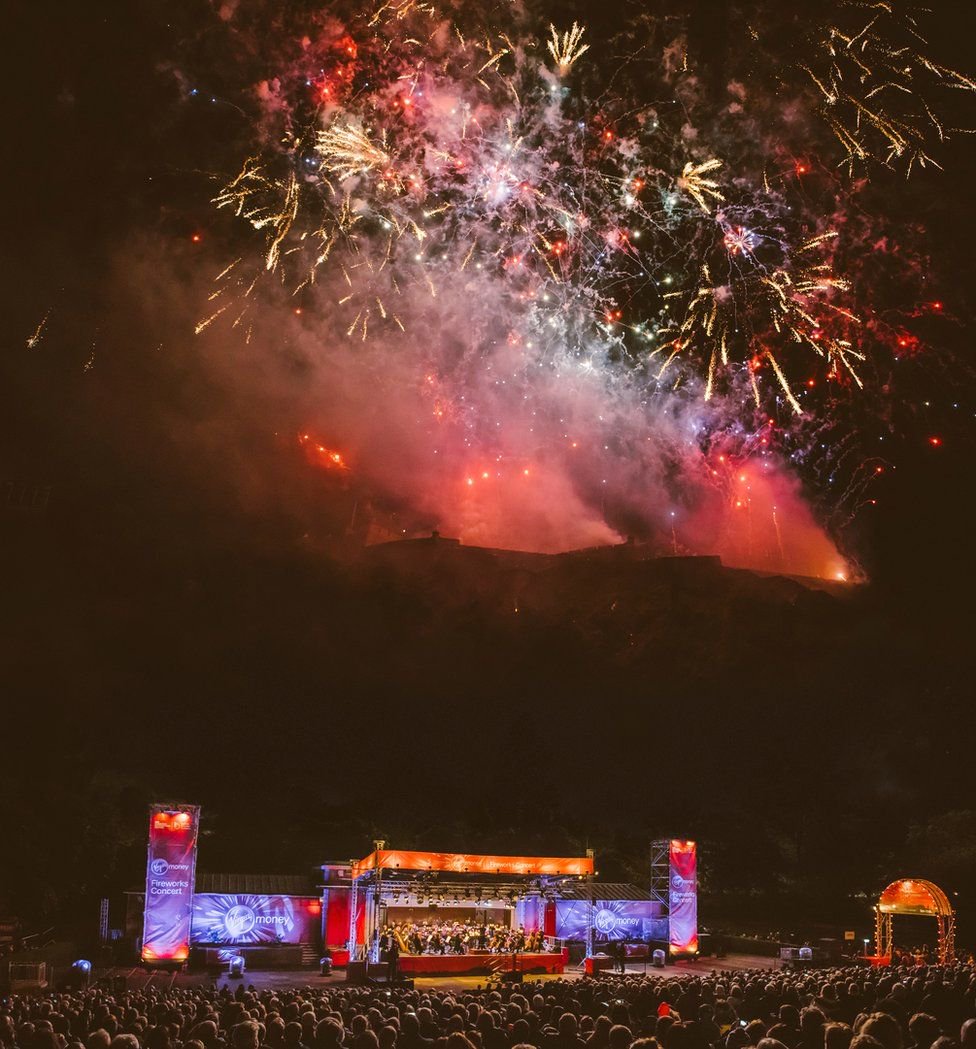}
	    & \includegraphics[height=0.20\columnwidth]{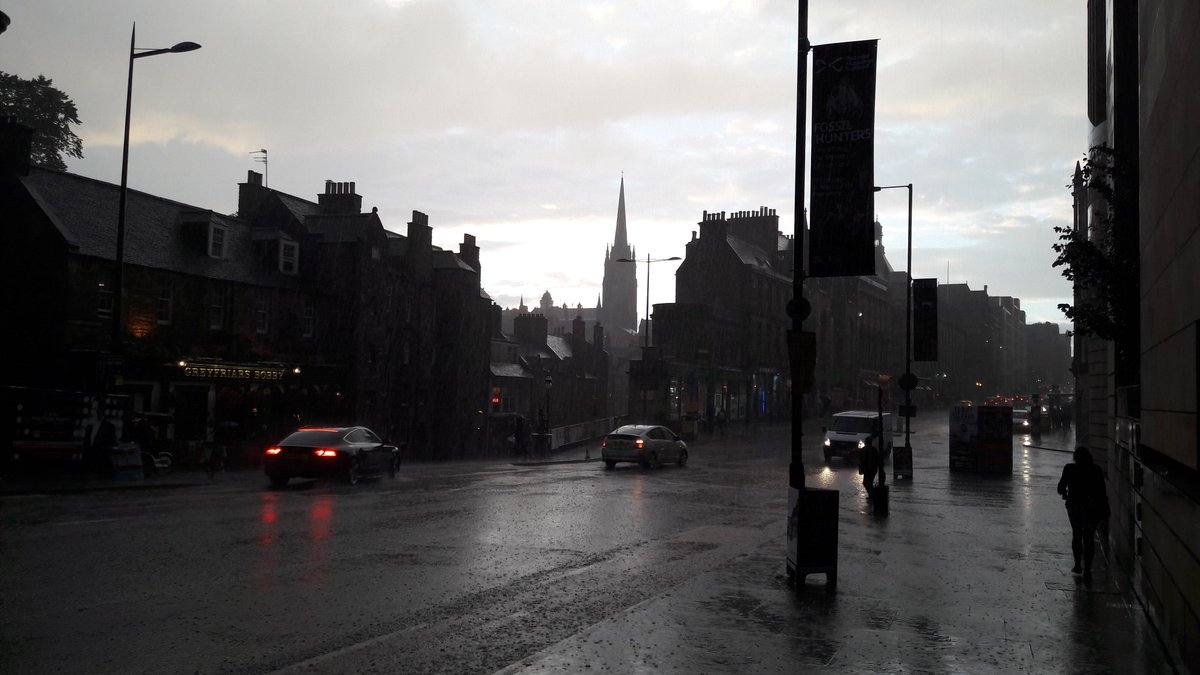}
		& \includegraphics[height=0.20\columnwidth]{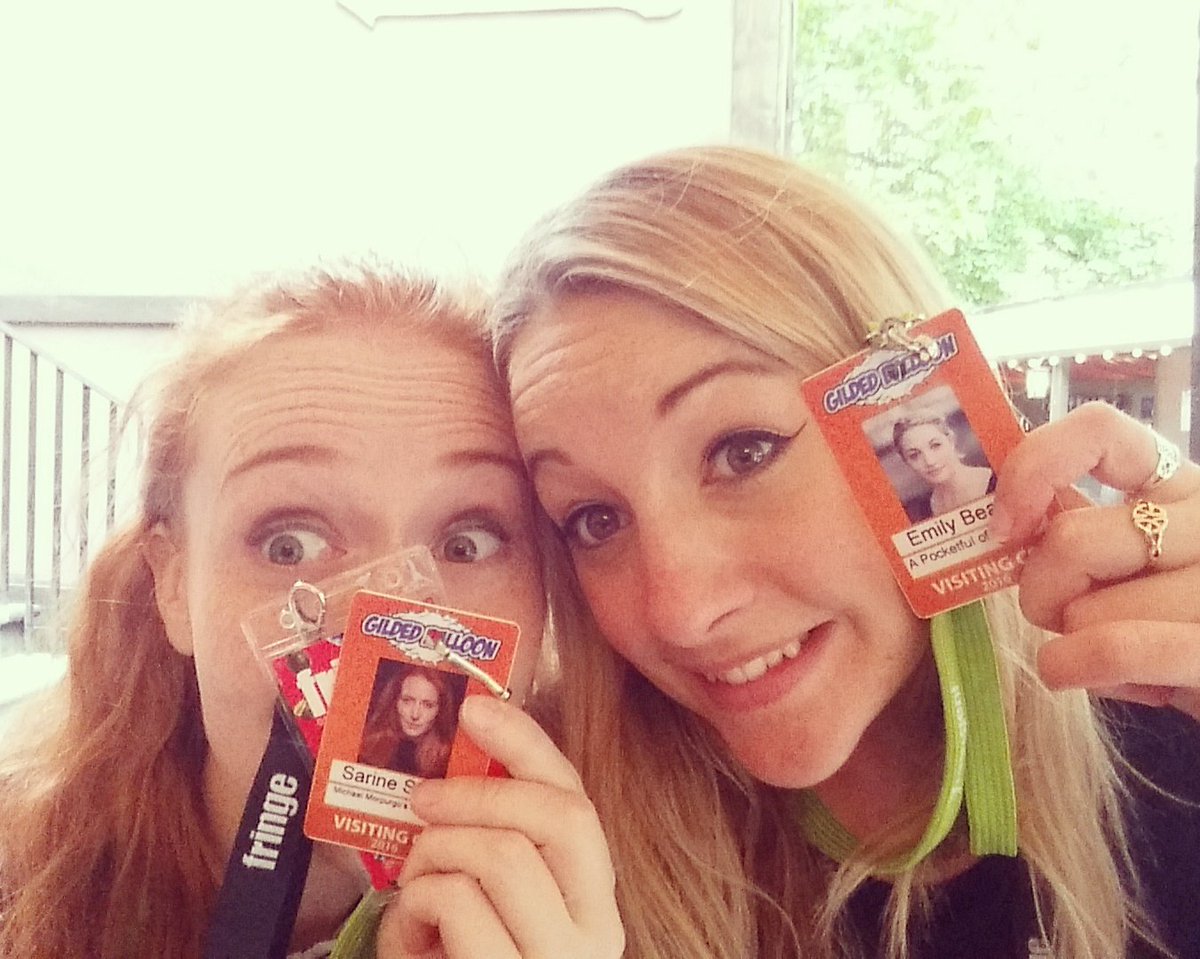}\\
        \midrule
		\small{\textbf{$GBT_{S}$}} 
		& \small{\#Duplicates$^\uparrow$, \#Retweets$^\uparrow$}
		& \small{\#Duplicates$^\uparrow$, \#Retweets$^\downarrow$}
		& \small{\#Retweets$^\uparrow$ \#Duplicates$^\downarrow$}
		& \small{\#Duplicates$^\downarrow$, \#Retweets$^\downarrow$}\\
	    & \includegraphics[height=0.20\columnwidth]{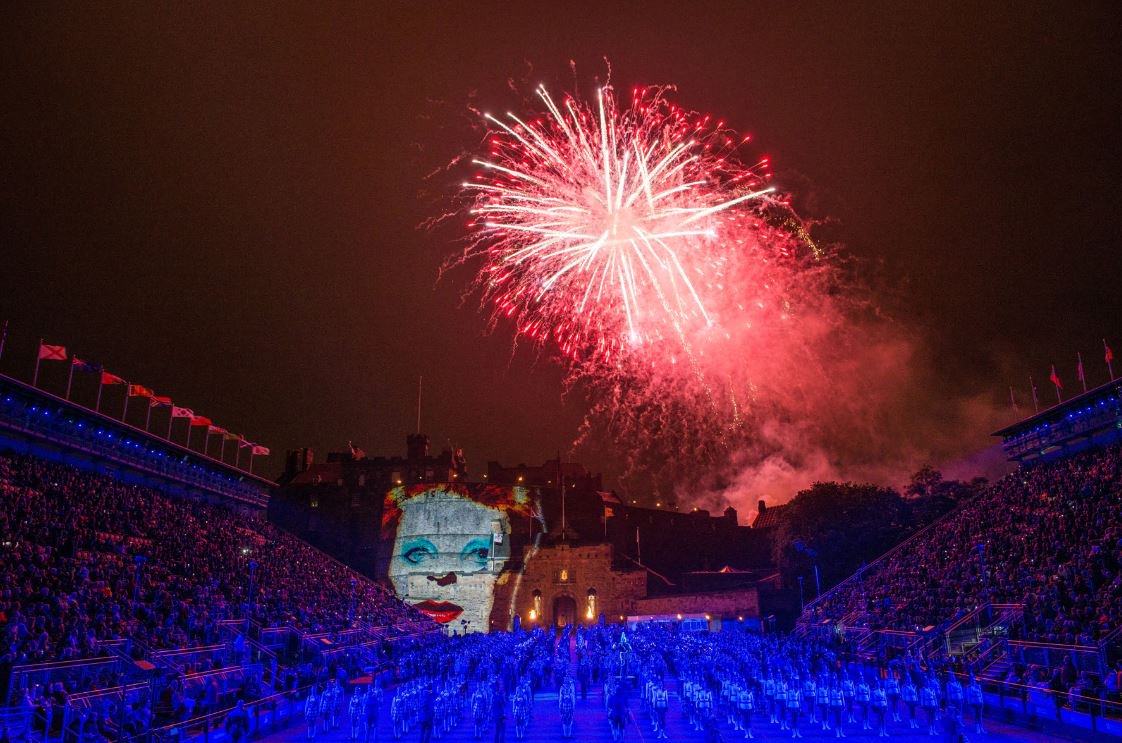}
	    & \includegraphics[height=0.20\columnwidth]{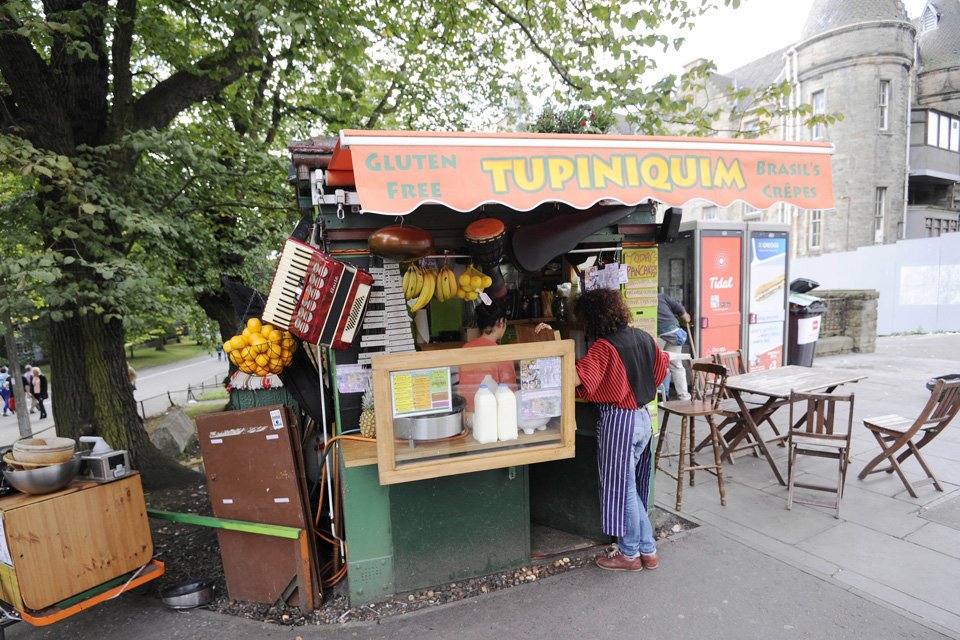}
	    & \includegraphics[height=0.20\columnwidth]{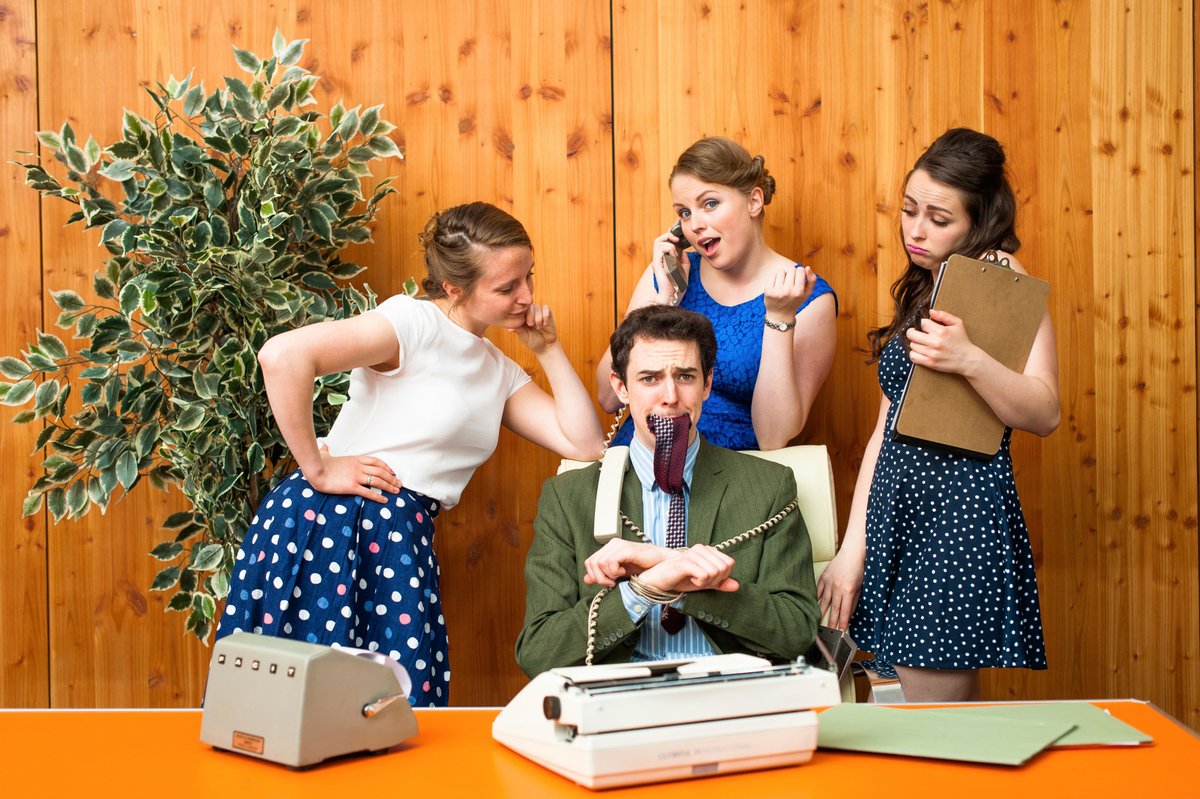}
        & \includegraphics[height=0.20\columnwidth]{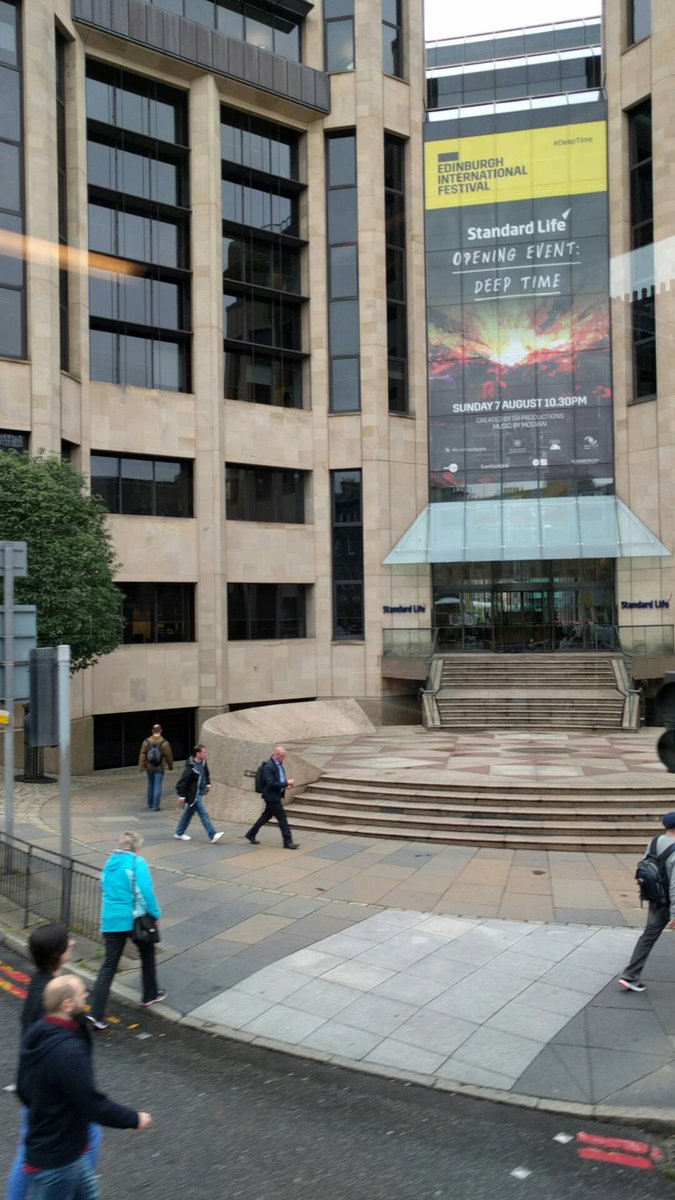}\\
		\midrule
		\small{\textbf{$GBT_{F}$}} 
		& \small{Visual$^\uparrow$, Social$^\uparrow$, Semantic$^\uparrow$}
		& \small{Semantic$^\uparrow$, Visual$^\uparrow$}
		& \small{Social$^\downarrow$, Semantic$^\uparrow$}
		& \small{Visual$^\downarrow$, Social$^\downarrow$, Semantic$^\downarrow$}\\
	    & \includegraphics[height=0.20\columnwidth]{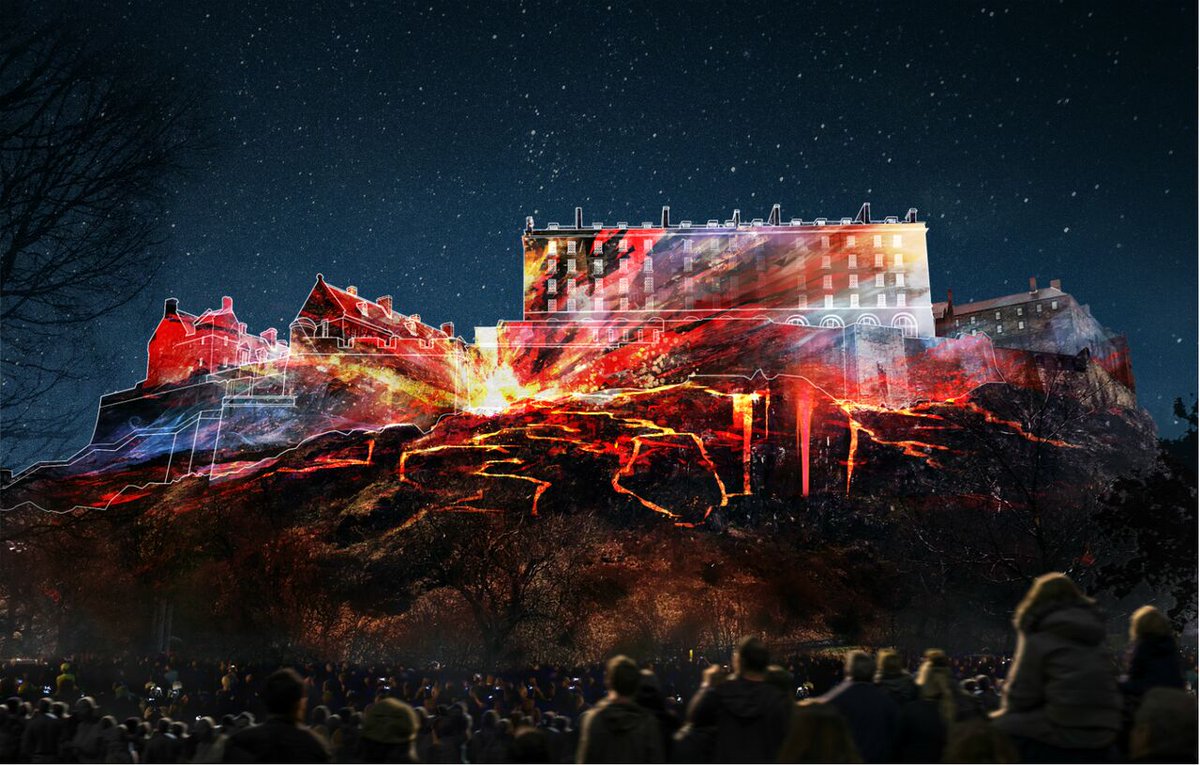}
	    & \includegraphics[height=0.20\columnwidth]{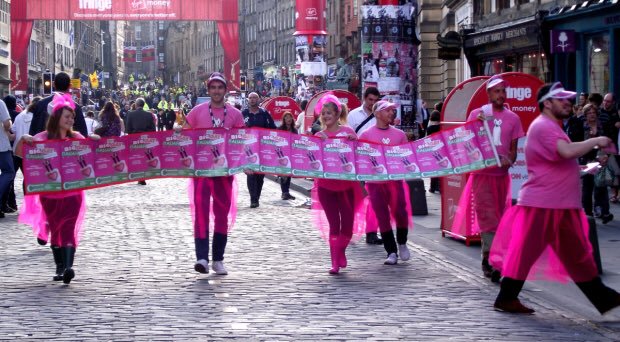}
	    & \includegraphics[height=0.20\columnwidth]{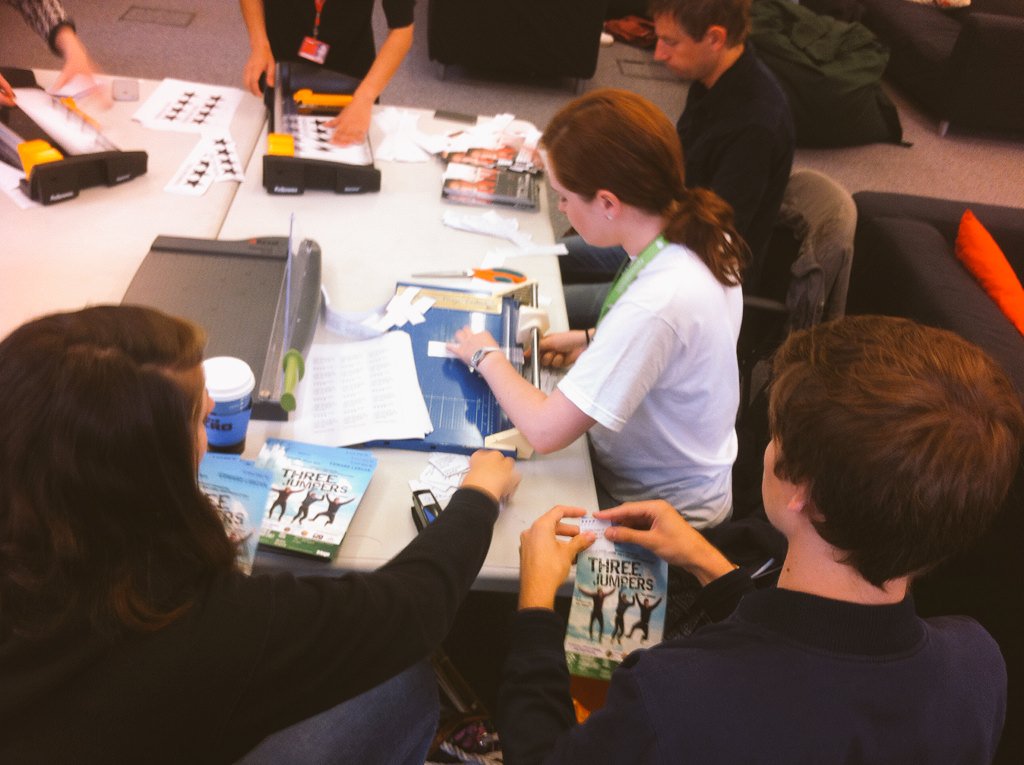}
        & \includegraphics[height=0.20\columnwidth]{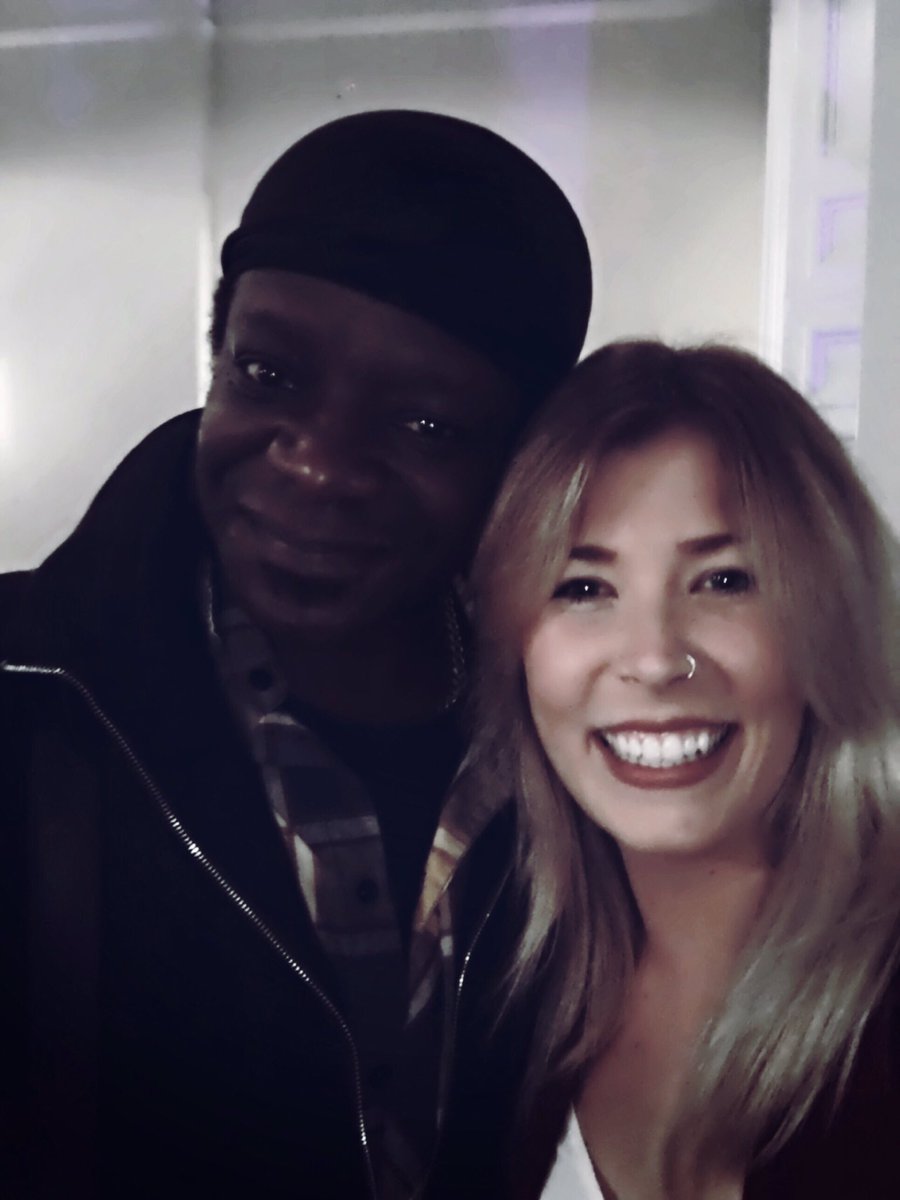}\\
	    \bottomrule
	\end{tabular}
	\caption{Examples of images ranked by four distinct models with increasing ranks from left to right.}
	\label{fig:results_example}
	\vspace{-5mm}
\end{table*}

\subsubsection{Filtering low quality photos.}
To detect synthetic and captioned images, we trained a logistic regression classifier (with L1 penalty and $\lambda = 1.0$). We applied 5-fold cross-validation and trained the model with the dataset from \cite{Wang2006}.
Features include luminance, edge histograms (with edges grouped into vertical, horizontal, 45 and 135 degrees orientations), dominant colors and corners ratio. After normalizing the size of all images to 240px width and 180px height we tuned different feature parameters. Namely, the domaint colors threshold was set to 600 and the ratio between the number of corners in synthetic/natural images was set to 0.1801, for a maximum distance of 20\%. 

Having extracted multiple featues, we experimented different combinations. Using luminance only, we managed to achieve a decent performance. We hypothesized that dominant colors and the ratio of pixels with the most common color (C1) were also good features that individually could distinguish both classes. Table~\ref{tab:synth_classifier_performance} shows the performance of both these features, which corresponds to our expectation. The C1 feature has an exceptionally good performance for a unidimensional classifier.

Selecting the best set of features, we reach 97\% precision on the NIPC dataset. 
However, as discussed previously, it seems that some of the images found on Twitter fall outside the categories defined by the authors of NPIC and thus we had to retrain the model. 
The classifier was retrained on the NIPC-Twitter dataset, making it now able to correctly classify as SPAM partially \textit{synthetic} images, that have, for instance, a photography as background (see Figure~\ref{fig:spam}). In the end, the final model trained on the NIPC-Twitter dataset achieved 91\% precision on the NPIC dataset.

\section{Conclusions}
We proposed a framework designed help news editors by ranking and filtering images extracted from social media, according to their quality in the context of news media. The framework takes advantage of visual, social and semantic feature groups. 
Using several methods of evaluation we show the importance of having these groups of features tackle the different problems associated with the task: 
\begin{itemize}
    \item Social features can be used as proxies to measure the interestingness and quality of an image but the lack of strong social signals does not directly imply the image is not news worthy. 
    
    \item Semantic features can be used to discard images that are generally not employed in the context of news media, such as \textit{selfies}, while giving priority to topics covered more often in the news. However, semantic features not only do not ensure the visual quality of images but also may not be of great help with images that have rare concepts associated with them, that the model was not able to interact with in the training phase. 
    
    \item Finally, visual features can be used to ensure the visual quality of an image but are not enough to ensure the interestingness and quality of the information it provides.
\end{itemize}

Consequently, the model that performed systematically better during evaluation was the one that leverages simultaneously these three feature groups

Finally, the above results were only possible to achieve in real-world social-media data because we deployed a thorough visual SPAM and redundancy filtering process. SPAM is a big part of the social-media, thus, we combined synthetic image detectors, captioned image filters, near-duplicate removal and other heuristics to clean low-quality data, allowing the ranking and filtering models to work with cleaner data.

\vspace{3mm}
\noindent
\textbf{Acknowledgements.} This work has been partially funded by the CMU Portugal research project GoLocal Ref. CMUP-ERI/TIC/0033/2014, by the H2020 ICT project COGNITUS with the grant agreement No 687605 and by the project NOVA LINCS Ref. UID/CEC/04516/2013.

\balance
\newcommand{\shownote}[1]{\unskip}
\newcommand{\showDOI}[1]{\unskip}
\newcommand{\showURL}[1]{\unskip}


\begin{thebibliography}{24}


\ifx \showCODEN    \undefined \def \showCODEN     #1{\unskip}     \fi
\ifx \showDOI      \undefined \def \showDOI       #1{#1}\fi
\ifx \showISBNx    \undefined \def \showISBNx     #1{\unskip}     \fi
\ifx \showISBNxiii \undefined \def \showISBNxiii  #1{\unskip}     \fi
\ifx \showISSN     \undefined \def \showISSN      #1{\unskip}     \fi
\ifx \showLCCN     \undefined \def \showLCCN      #1{\unskip}     \fi
\ifx \shownote     \undefined \def \shownote      #1{#1}          \fi
\ifx \showarticletitle \undefined \def \showarticletitle #1{#1}   \fi
\ifx \showURL      \undefined \def \showURL       {\relax}        \fi
\providecommand\bibfield[2]{#2}
\providecommand\bibinfo[2]{#2}
\providecommand\natexlab[1]{#1}
\providecommand\showeprint[2][]{arXiv:#2}

\bibitem[\protect\citeauthoryear{Arapakis, Peleja, Cambazoglu, and
  Magalhaes}{Arapakis et~al\mbox{.}}{2016}]%
        {arapakis2016linguistic}
\bibfield{author}{\bibinfo{person}{Ioannis Arapakis}, \bibinfo{person}{Filipa
  Peleja}, \bibinfo{person}{Berkant~Barla Cambazoglu}, {and}
  \bibinfo{person}{Joao Magalhaes}.} \bibinfo{year}{2016}\natexlab{}.
\newblock \showarticletitle{Linguistic Benchmarks of Online News Article
  Quality.}. In \bibinfo{booktitle}{{\em ACL (1)}}.
\newblock


\bibitem[\protect\citeauthoryear{Athitsos, Swain, and Frankel}{Athitsos
  et~al\mbox{.}}{1997}]%
        {athitsos1997distinguishing}
\bibfield{author}{\bibinfo{person}{Vassilis Athitsos},
  \bibinfo{person}{Michael~J Swain}, {and} \bibinfo{person}{Charles Frankel}.}
  \bibinfo{year}{1997}\natexlab{}.
\newblock \showarticletitle{Distinguishing photographs and graphics on the
  world wide web}. In \bibinfo{booktitle}{{\em Content-Based Access of Image
  and Video Libraries, 1997. Proceedings. IEEE Workshop on}}. IEEE,
  \bibinfo{pages}{10--17}.
\newblock


\bibitem[\protect\citeauthoryear{Chen and Guestrin}{Chen and Guestrin}{2016}]%
        {chen2016xgboost}
\bibfield{author}{\bibinfo{person}{Tianqi Chen} {and} \bibinfo{person}{Carlos
  Guestrin}.} \bibinfo{year}{2016}\natexlab{}.
\newblock \showarticletitle{Xgboost: A scalable tree boosting system}. In
  \bibinfo{booktitle}{{\em Proceedings of the 22nd acm sigkdd international
  conference on knowledge discovery and data mining}}. ACM,
  \bibinfo{pages}{785--794}.
\newblock


\bibitem[\protect\citeauthoryear{Dhar, Ordonez, and Berg}{Dhar
  et~al\mbox{.}}{2011}]%
        {dhar2011high}
\bibfield{author}{\bibinfo{person}{Sagnik Dhar}, \bibinfo{person}{Vicente
  Ordonez}, {and} \bibinfo{person}{Tamara~L Berg}.}
  \bibinfo{year}{2011}\natexlab{}.
\newblock \showarticletitle{High level describable attributes for predicting
  aesthetics and interestingness}. In \bibinfo{booktitle}{{\em Computer Vision
  and Pattern Recognition (CVPR), 2011 IEEE Conference on}}. IEEE.
\newblock


\bibitem[\protect\citeauthoryear{Freeman et~al\mbox{.}}{Freeman
  et~al\mbox{.}}{2007}]%
        {freeman2007photographer}
\bibfield{author}{\bibinfo{person}{Michael Freeman} {et~al\mbox{.}}}
  \bibinfo{year}{2007}\natexlab{}.
\newblock \bibinfo{booktitle}{{\em The Photographer's Eye: Composition and
  Design for Better Digital Photos}}.
\newblock \bibinfo{publisher}{CRC Press}.
\newblock


\bibitem[\protect\citeauthoryear{Friedman, Hastie, and Tibshirani}{Friedman
  et~al\mbox{.}}{2001}]%
        {friedman2001elements}
\bibfield{author}{\bibinfo{person}{Jerome Friedman}, \bibinfo{person}{Trevor
  Hastie}, {and} \bibinfo{person}{Robert Tibshirani}.}
  \bibinfo{year}{2001}\natexlab{}.
\newblock \bibinfo{booktitle}{{\em The elements of statistical learning}}.
\newblock \bibinfo{publisher}{Springer series in statistics New York}.
\newblock


\bibitem[\protect\citeauthoryear{Harper, Raban, Rafaeli, and Konstan}{Harper
  et~al\mbox{.}}{2008}]%
        {harper2008predictors}
\bibfield{author}{\bibinfo{person}{F~Maxwell Harper}, \bibinfo{person}{Daphne
  Raban}, \bibinfo{person}{Sheizaf Rafaeli}, {and} \bibinfo{person}{Joseph~A
  Konstan}.} \bibinfo{year}{2008}\natexlab{}.
\newblock \showarticletitle{Predictors of answer quality in online Q\&A sites}.
  In \bibinfo{booktitle}{{\em Proceedings of the SIGCHI Conference on Human
  Factors in Computing Systems}}. ACM.
\newblock


\bibitem[\protect\citeauthoryear{Hasler and S{\"u}sstrunk}{Hasler and
  S{\"u}sstrunk}{2003}]%
        {hasler2003measuring}
\bibfield{author}{\bibinfo{person}{David Hasler} {and} \bibinfo{person}{Sabine
  S{\"u}sstrunk}.} \bibinfo{year}{2003}\natexlab{}.
\newblock \showarticletitle{Measuring colourfulness in natural images}. In
  \bibinfo{booktitle}{{\em Proc. IST/SPIE Electronic Imaging 2003: Human Vision
  and Electronic Imaging VIII}}, Vol.~\bibinfo{volume}{5007}.
\newblock


\bibitem[\protect\citeauthoryear{Isola, Parikh, Torralba, and Oliva}{Isola
  et~al\mbox{.}}{2011}]%
        {isola2011understanding}
\bibfield{author}{\bibinfo{person}{Phillip Isola}, \bibinfo{person}{Devi
  Parikh}, \bibinfo{person}{Antonio Torralba}, {and} \bibinfo{person}{Aude
  Oliva}.} \bibinfo{year}{2011}\natexlab{}.
\newblock \showarticletitle{Understanding the intrinsic memorability of
  images}. In \bibinfo{booktitle}{{\em Advances in Neural Information
  Processing Systems}}.
\newblock


\bibitem[\protect\citeauthoryear{Joachims}{Joachims}{2006}]%
        {joachims2006training}
\bibfield{author}{\bibinfo{person}{Thorsten Joachims}.}
  \bibinfo{year}{2006}\natexlab{}.
\newblock \showarticletitle{Training linear SVMs in linear time}. In
  \bibinfo{booktitle}{{\em Proceedings of the 12th ACM SIGKDD international
  conference on Knowledge discovery and data mining}}. ACM.
\newblock


\bibitem[\protect\citeauthoryear{Kobre and Brill}{Kobre and Brill}{2004}]%
        {kobre2004photojournalism}
\bibfield{author}{\bibinfo{person}{Kenneth Kobre} {and} \bibinfo{person}{Betsy
  Brill}.} \bibinfo{year}{2004}\natexlab{}.
\newblock \bibinfo{booktitle}{{\em Photojournalism: the professionals'
  approach}}. Vol.~\bibinfo{volume}{5}.
\newblock \bibinfo{publisher}{Focal Press Burlington, MA}.
\newblock


\bibitem[\protect\citeauthoryear{Krawetz}{Krawetz}{[n. d.]}]%
        {HackerFactor_pHash}
\bibfield{author}{\bibinfo{person}{Neal Krawetz}.} \bibinfo{year}{[n.
  d.]}\natexlab{}.
\newblock \bibinfo{title}{Looks Like It}.
\newblock
  \bibinfo{howpublished}{\url{http://www.hackerfactor.com/blog/index.php?/archives/432-Looks-Like-It.html}}.
    (\bibinfo{year}{[n. d.]}).
\newblock
\newblock
\shownote{Accessed: 2017-03-23.}


\bibitem[\protect\citeauthoryear{Kwak, Lee, Park, and Moon}{Kwak
  et~al\mbox{.}}{2010}]%
        {Kwak:2010:TSN:1772690.1772751}
\bibfield{author}{\bibinfo{person}{Haewoon Kwak}, \bibinfo{person}{Changhyun
  Lee}, \bibinfo{person}{Hosung Park}, {and} \bibinfo{person}{Sue Moon}.}
  \bibinfo{year}{2010}\natexlab{}.
\newblock \showarticletitle{What is Twitter, a Social Network or a News
  Media?}. In \bibinfo{booktitle}{{\em Proceedings of the 19th International
  Conference on World Wide Web}} {\em (\bibinfo{series}{WWW '10})}.
  \bibinfo{publisher}{ACM}, \bibinfo{address}{New York, NY, USA}.
\newblock
\showISBNx{978-1-60558-799-8}
\showDOI{%
\url{https://doi.org/10.1145/1772690.1772751}}


\bibitem[\protect\citeauthoryear{Lienhart and Hartmann}{Lienhart and
  Hartmann}{2002}]%
        {Lienhart2002}
\bibfield{author}{\bibinfo{person}{R Lienhart} {and} \bibinfo{person}{A
  Hartmann}.} \bibinfo{year}{2002}\natexlab{}.
\newblock \showarticletitle{{Classifying images on the web automatically}}.
\newblock \bibinfo{journal}{{\em Journal of Electronic Imaging\/}}
  \bibinfo{volume}{11}, \bibinfo{number}{4} (\bibinfo{year}{2002}),
  \bibinfo{pages}{445--454}.
\newblock
\showISSN{10179909 (ISSN)}
\showDOI{%
\url{https://doi.org/10.1117/1.1502259}}


\bibitem[\protect\citeauthoryear{Liu and Huet}{Liu and Huet}{2013}]%
        {liu2013eventenricher}
\bibfield{author}{\bibinfo{person}{Xueliang Liu} {and} \bibinfo{person}{Benoit
  Huet}.} \bibinfo{year}{2013}\natexlab{}.
\newblock \showarticletitle{EventEnricher: a novel way to collect media
  illustrating events}. In \bibinfo{booktitle}{{\em Proceedings of the 3rd ACM
  conference on International conference on multimedia retrieval}}. ACM.
\newblock


\bibitem[\protect\citeauthoryear{Marchesotti, Perronnin, Larlus, and
  Csurka}{Marchesotti et~al\mbox{.}}{2011}]%
        {marchesotti2011assessing}
\bibfield{author}{\bibinfo{person}{Luca Marchesotti}, \bibinfo{person}{Florent
  Perronnin}, \bibinfo{person}{Diane Larlus}, {and} \bibinfo{person}{Gabriela
  Csurka}.} \bibinfo{year}{2011}\natexlab{}.
\newblock \showarticletitle{Assessing the aesthetic quality of photographs
  using generic image descriptors}. In \bibinfo{booktitle}{{\em Computer Vision
  (ICCV), 2011 IEEE International Conference on}}. IEEE.
\newblock


\bibitem[\protect\citeauthoryear{Martins and Correia}{Martins and
  Correia}{2017}]%
        {martins2017semi}
\bibfield{author}{\bibinfo{person}{Pedro Martins} {and} \bibinfo{person}{Nuno
  Correia}.} \bibinfo{year}{2017}\natexlab{}.
\newblock \showarticletitle{Semi-automatic Video Assessment System}. In
  \bibinfo{booktitle}{{\em Proceedings of the 15th International Workshop on
  Content-Based Multimedia Indexing}}. ACM, \bibinfo{pages}{33}.
\newblock


\bibitem[\protect\citeauthoryear{McMinn, Moshfeghi, and Jose}{McMinn
  et~al\mbox{.}}{2013}]%
        {mcminn2013building}
\bibfield{author}{\bibinfo{person}{Andrew~J McMinn}, \bibinfo{person}{Yashar
  Moshfeghi}, {and} \bibinfo{person}{Joemon~M Jose}.}
  \bibinfo{year}{2013}\natexlab{}.
\newblock \showarticletitle{Building a large-scale corpus for evaluating event
  detection on twitter}. In \bibinfo{booktitle}{{\em Proceedings of the 22nd
  ACM international conference on Information \& Knowledge Management}}. ACM.
\newblock


\bibitem[\protect\citeauthoryear{Mcparlane, Mcminn, and Jose}{Mcparlane
  et~al\mbox{.}}{2014}]%
        {Mcparlane2014}
\bibfield{author}{\bibinfo{person}{Philip~J Mcparlane},
  \bibinfo{person}{Andrew~J Mcminn}, {and} \bibinfo{person}{Joemon~M Jose}.}
  \bibinfo{year}{2014}\natexlab{}.
\newblock \showarticletitle{{" Picture the scene ..."; Visually Summarising
  Social Media Events Categories and Subject Descriptors}}.
\newblock \bibinfo{journal}{{\em Proceedings of the 23rd ACM International
  Conference on Conference on Information and Knowledge Management\/}}
  (\bibinfo{year}{2014}), \bibinfo{pages}{1459--1468}.
\newblock
\showISBNx{9781450325981}
\showDOI{%
\url{https://doi.org/10.1145/2661829.2661923}}


\bibitem[\protect\citeauthoryear{Murphy}{Murphy}{2012}]%
        {murphy2012machine}
\bibfield{author}{\bibinfo{person}{Kevin~P Murphy}.}
  \bibinfo{year}{2012}\natexlab{}.
\newblock \bibinfo{booktitle}{{\em Machine learning: a probabilistic
  perspective}}.
\newblock \bibinfo{publisher}{MIT press}.
\newblock


\bibitem[\protect\citeauthoryear{Schinas, Papadopoulos, Kompatsiaris, and
  Mitkas}{Schinas et~al\mbox{.}}{2015}]%
        {Schinas2015}
\bibfield{author}{\bibinfo{person}{Manos Schinas}, \bibinfo{person}{Symeon
  Papadopoulos}, \bibinfo{person}{Yiannis Kompatsiaris}, {and}
  \bibinfo{person}{Pericles~A. Mitkas}.} \bibinfo{year}{2015}\natexlab{}.
\newblock \showarticletitle{{Visual Event Summarization on Social Media using
  Topic Modelling and Graph-based Ranking Algorithms}}. In
  \bibinfo{booktitle}{{\em Proceedings of the 5th ACM on International
  Conference on Multimedia Retrieval - ICMR '15}}. \bibinfo{publisher}{ACM
  Press}, \bibinfo{address}{New York, New York, USA},
  \bibinfo{pages}{203--210}.
\newblock
\showISBNx{9781450332743}
\showDOI{%
\url{https://doi.org/10.1145/2671188.2749407}}


\bibitem[\protect\citeauthoryear{Smith}{Smith}{2007}]%
        {smith2007overview}
\bibfield{author}{\bibinfo{person}{Ray Smith}.}
  \bibinfo{year}{2007}\natexlab{}.
\newblock \showarticletitle{An overview of the Tesseract OCR engine}. In
  \bibinfo{booktitle}{{\em Document Analysis and Recognition, 2007. ICDAR 2007.
  Ninth International Conference on}}, Vol.~\bibinfo{volume}{2}. IEEE.
\newblock


\bibitem[\protect\citeauthoryear{Tang, Dai, and Zhang}{Tang
  et~al\mbox{.}}{2012}]%
        {Tang2012}
\bibfield{author}{\bibinfo{person}{Z Tang}, \bibinfo{person}{Y Dai}, {and}
  \bibinfo{person}{X Zhang}.} \bibinfo{year}{2012}\natexlab{}.
\newblock \showarticletitle{{Perceptual hashing for color images using
  invariant moments}}.
\newblock \bibinfo{journal}{{\em Appl. Math\/}} (\bibinfo{year}{2012}).
\newblock
\showURL{%
\url{http://www.naturalspublishing.com/files/published/54515x71g3omq1.pdf}}


\bibitem[\protect\citeauthoryear{Wang and Kan}{Wang and Kan}{2006}]%
        {Wang2006}
\bibfield{author}{\bibinfo{person}{Fei Wang} {and} \bibinfo{person}{Min-yen
  Kan}.} \bibinfo{year}{2006}\natexlab{}.
\newblock \showarticletitle{{NPIC : Hierarchical Synthetic Image Classification
  Using Image Search and Generic Features}}.
\newblock \bibinfo{journal}{{\em Work\/}} (\bibinfo{year}{2006}),
  \bibinfo{pages}{473--482}.
\newblock
\showISBNx{3540360182}
\showISSN{16113349}


\end{thebibliography}
\end{document}